\documentclass[a4paper,superscriptaddress,showpacs,preprint,prl,floatfix]{revtex4} 
\usepackage{tabularx}
\usepackage{amsmath}
\usepackage{amssymb}
\usepackage{array}
\usepackage{multirow}
\usepackage{rotating}
\usepackage{subfigure}
\usepackage{graphicx}
\begin{document}
\setlength{\tabcolsep}{1.5mm} 
\setcounter{totalnumber}{4}
\setcounter{topnumber}{4}
\setlength{\voffset}{-0cm}
\setlength{\hoffset}{-0.cm}
\addtolength{\textheight}{1.1cm}




\newcommand{\figwidth}{0.90\columnwidth}
\newcommand{\eq}[1]{Eq.(\ref{#1})}
\newcommand{\fig}[1]{Fig.~\ref{#1}}
\newcommand{\sect}[1]{Sec.~\ref{#1}}
\newcommand{\avg}[1]{{\langle #1 \rangle}}
\newcommand{\olcite}[1]{Ref.~\onlinecite{#1}}




\title[De Novo protein folding and design]{Transferable coarse-grained potential for \textit{de novo} protein folding and design}
\author{Ivan Coluzza}
\email{ivan.coluzza@univie.ac.at}
\affiliation{Faculty of Physics, University of Vienna, Boltzmanngasse 5, 1090 Vienna, Austria}
\pacs{ 87.15.Cc, 05.20.-y, 87.10.Rt}
\begin{abstract}
Protein folding and design are major biophysical problems, the solution of which would lead to important applications especially in medicine. Here a novel protein model capable of simultaneously provide quantitative protein design and folding is introduced. With computer simulations it is shown that, for a large set of real protein structures, the model produces designed sequences with similar physical properties to the corresponding natural occurring sequences. The designed sequences are not yet fully realistic and require further experimental testing. For an independent set of proteins, notoriously difficult to fold,  the correct folding of both the designed and the natural sequences is also demonstrated. The folding properties are characterized by free energy calculations. which not only are consistent among natural and designed proteins, but we also show a remarkable precision when the folded structures are compared to the experimentally determined ones. Ultimately, this novel coarse-grained protein model is unique in the combination of its fundamental three features: its simplicity, its ability to produce natural foldable designed sequences, and its structure prediction precision. The latter demonstrated by free energy calculations. It is also remarkable that low frustration sequences can be obtained with such a simple and universal design procedure, and that the folding of natural proteins shows funnelled free energy landscapes without the need of any potentials based on the native structure.
\end{abstract}
\maketitle
Computer simulations of the protein folding process have in the last ten years reached amazing level of description and accuracy~\cite{VoeglerSmith2001,Pande2003,Hamelberg2004,Seibert2005,Tozzini2005a,Henzler-Wildman2007a,Mu2007,Huang2008,Laio2008,Bereau2009a,Shaw2010a,Zhang2010,Ikebe2011,Lindorff-Larsen2011,Biology2012,Kapoor2013}. The power of the computers and the understanding of the physics that governs folding allows now for a large screening of the experimental data for instance collected in the Protein Data Bank~\cite{Berman2000}.
From a theoretical point of view a successful approach is  the``minimal frustration principle'' (MFP) ~\cite{BRYNGELSON1987a,Shakhnovich1993a,Wolynes2012} in which protein folding is described as a downhill sliding process in a low frustration energy landscape (``funnelled'' shaped) towards the native state.  While MFP has been proven for lattice heteropolymers~\cite{Shakhnovich1993a,Gutin1993a,Shakhnovich1994,Shakhnovich1990,Coluzza2003,Coluzza2004,Coluzza2007a,Abeln2008}, in more realistic protein representations a residual frustration which prevents the systematic prediction of the native structure of natural sequences is often observed. Off-lattice instead MFP is used as a main justification for the use of structure-based potentials such as the GO~\cite{Go1983} and elastic models~\cite{Atilgan2001}. In fact, there is still space for development of transferable models that are capable of systematic associating the experimentally determined native structure to natural sequence. Surprisingly, with the exception of few notable examples~\cite{Dahiyat1997,Desjarlais1995,Desjarlais1999,Hellinga1991,DiStasio2012,Rothlisberger2008,Dahiyat1997a,Kuhlman2003,Koga2012}, it has also been extremely difficult to artificially construct sequences capable of folding into given target protein structures. 
The group of David Baker~\cite{Koga2012} introduced a novel procedure to select sequences with low frustration capable of correctly refolding in vitro to their target structure with a success rate between 8\% and up to 40\% of the total trials. In their work the authors have introduced a set of rules for the design of the local amino acids interactions   to disfavour non-native states. After many iterations, a refolding calculation filters out about 90\% of the initial sequences that  are found not to have  a funnelled energy landscape. The complexity of Baker's  procedure demonstrates that is not easy to produce sequences with low frustration.

Here we present a novel protein model where low frustration folding is observed both for  natural and designed sequences, the latter  obtained without the need of negative design. The novel model is obtained from the optimization of the residue-residue and residue-solvent interaction energy terms under the condition that a large number of sequences designed for 125 test proteins are equal to the corresponding natural sequences. As a result, designed sequences with our model are for several properties comparable to natural ones and fold with a low frustration free energy landscape. We additionally demonstrated that for 15 additional randomly selected proteins, notoriously difficult to fold~\cite{Tsai2003,Kinch2011}, the natural sequences correctly refolded to their corresponding native structures with a remarkable precision between $2.5$ and $5$~\AA. In other words both quantitative protein design and folding are possible simultaneously. We anticipate that our methodology will have direct application for protein design and structure prediction, but also we expect that it will become a reference point for the development of alternative protein models. For instance, a more or less accurate description can be obtained by adding or removing details from our model, under the condition that the minimum constraint principle remains satisfied.

\begin{figure}[ht]
\includegraphics[width=1.00\columnwidth]{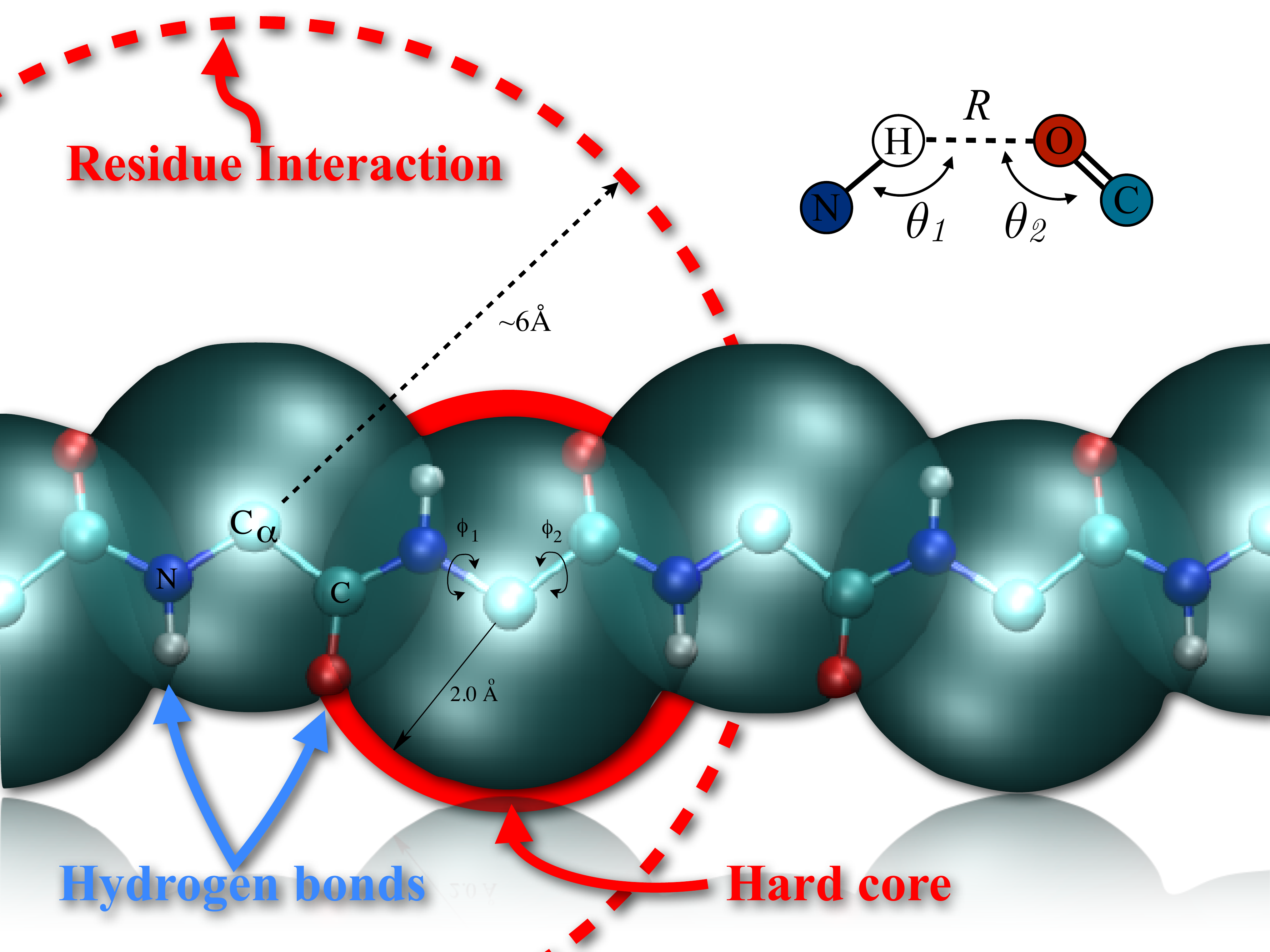}\caption{Real-space representation of the backbone of the Caterpillar model. The large blue sphere represent the self-avoidance volume $R_{HC}=2.0~\text{\AA}$ of the $C_{\alpha}$ atoms, while the interaction radius of each residue is represented by the large dashed circle or radius 6~\AA (see Eq.~S2 in Methods). The H and O atoms interact through a 10-12 Lennard-Jones  potential tuned with a quadratic orientation term that selects for alignment of the C, H, O, and N atoms involved in a bond (see top right inset and Eq.~S1 in Methods). The backbone fluctuates only around the torsional angles $\phi_1$ and $\phi_2$.}\label{Model}
\end{figure}

\begin{figure}[ht]
\includegraphics[width=.70\columnwidth]{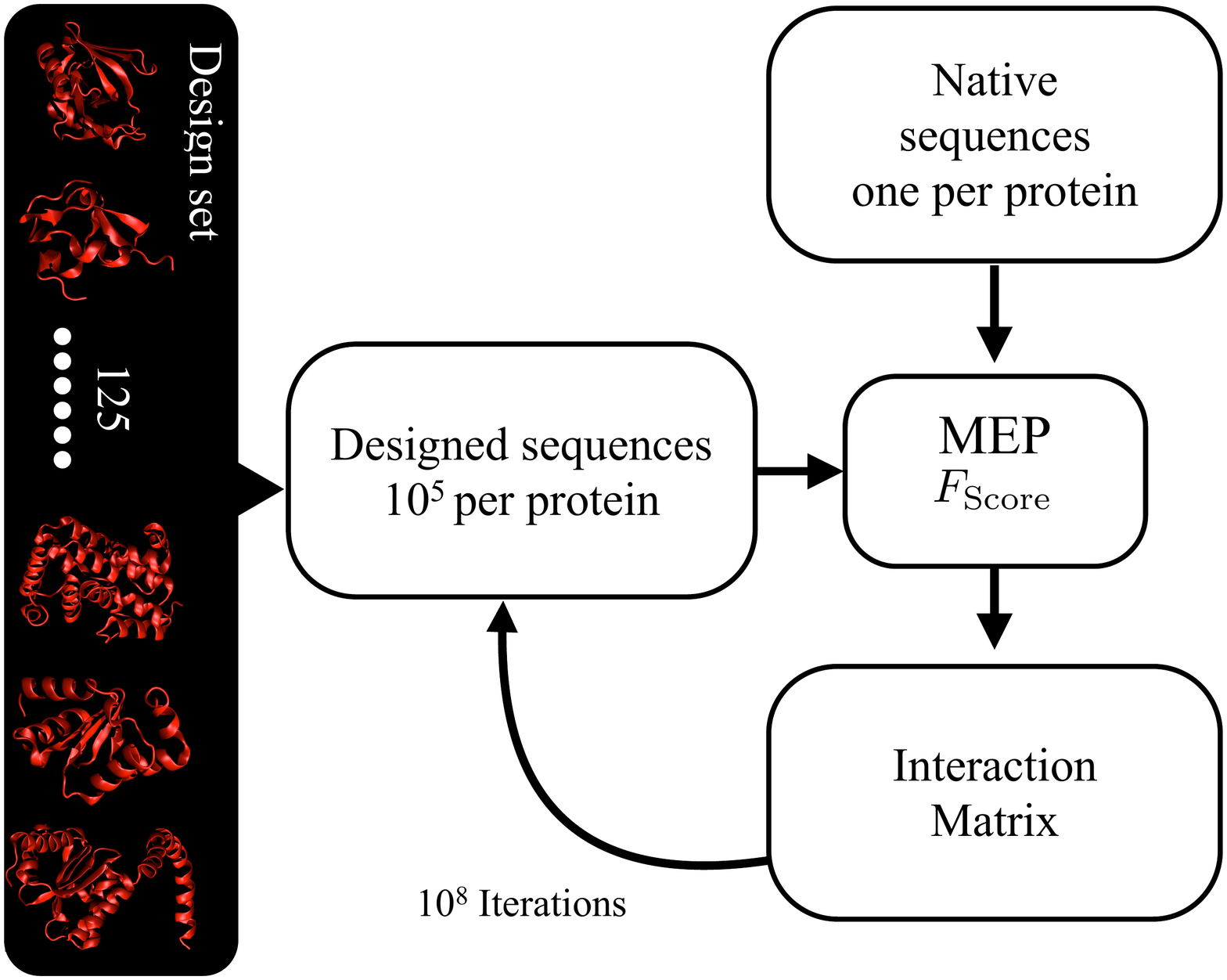}\caption{Schematic representation of the MEP algorithm. For a trial set of the $\epsilon$, $E_{\text{HOH}}$, and $\Omega$  parameters and for each protein in the training set a large number ($10^5$) of sequences with composition fixed to the natural one are generated following the design scheme in the SM. The scoring function $F_{\text{score}}$ (Eq.S16 in the SM) is then evaluated and the trial parameters are accepted or reject according to a Metropolis like scheme. New parameter sets are generated at each iteration, and the sequences of the proteins in the training set are re-designed by $10^5$ simple pair residue swapping moves, which are accepted or rejected according to a standard Metropolis algorithm with the energy defined in Eq.S4 (see SM). During each design iteration, the HP and energy profiles (Eq.S16 in the SM) are averaged over the observed sequences weighted by their Boltzmann weight. The averaging guarantees that the profiles are calculated over the most probable sequences that, as we showed previously~\cite{Coluzza2011}, are robust against mutations and are more thermally stable. After $\sim 10^8$ iterations the interaction parameters converged to their final values: $\Omega=21.0\pm0.5$ and $E_{\text{HOH}}=0.015\pm0.001$, and the residue-residue $\epsilon$ interaction parameters which are listed in Tab.~S1 of the SM }\label{MEP_Scheme}
\end{figure}

Recently we have presented many results that point to the existence of a "minimal constraint principle" (MCP)~\cite{Coluzza2012c,Coluzza2012b,Coluzza2011}, according to which for a heteropolymer to be designable and foldable it is sufficient that chain is decorated with directional interactions that constrain the configurational space. In the case of proteins we have shown (Caterpillar model~\cite{Coluzza2011}) that the minimum set of constraints translates into the combination of just the backbone molecular geometry and the backbone hydrogen bond interactions (see Fig.~\ref{Model}).  
In what follows we will show that by optimizing the interactions under the condition that natural and designed sequences are the same at constant amino acid composition for a large set of proteins, we will also quantitatively predict the folded structures of natural and designed sequences with similar accuracy. This is possible because our model includes the correct set of interactions that satisfy the MCP and, accordingly, the design procedure~\cite{Coluzza2011} alone is capable of predicting if a sequence, either natural or artificial, will fold to the target structure. Since we cannot model the particular evolutionary pressure that determined the natural amino acid composition, we chose to keep it constant. Such pressure could be due to many factors such as the particular function of the protein or the difficulty of synthesizing each amino acid type. The ansatz of this work is that folding and design can occur also outside such conditions and that is possible to design a foldable artificial protein from an infinite bath of amino acids. Hence, the above evolutionary pressure is taken into account by fixing the composition to the natural one. The optimization scheme that we used is  the maximum entropy principle (MEP) already tested for proteins by Seno et al.~\cite{Seno2008}. MEP states that the more information is used to model  a system the lower the associated entropy will be~\cite{Shannon1948}. Hence, in order to find the optimal parameters that require the least amount of information, all is needed is to maximize the entropy associated with the probability of observing a given protein $P(S_i,\Gamma_j)$, where $S_i$ indicates to the sequence and $\Gamma_j$ the three dimensional structure. The derivation follows closely the one used in the work of Seno et al.~\cite{Seno2008} (the full derivation is in the Supplemental Material~\footnote{See Supplemental Material for details about the model and simulations techniques together with the derivation of the scoring function with the Maximum Entropy Principle.} to save space) and we determined that the entropy maximum corresponds to  the values of the model parameters  ($\epsilon$, $E_{\text{HOH}}$, and $\Omega$ in  Eq.~\ref{Scoring_F}) at which the amino acid hydrophobic/hydrophilic (HP) profile~\cite{Dolittle1989} and the interaction energy of each residue with all other are simultaneously equal to the natural ones:
\begin{eqnarray}
  F_{\text{score}}=\sum_j^{N_{\text{Prot}}}  \sum_k^{N_j} \left (\sum_i^{N_{\text{Seq}}} P(S_i,\Gamma_j) E_{\text{Sol}}^{ik} - E_{\text{Sol}}^{\text{Real}_{jk}}\right)^2 + \nonumber\\ \sum_j^{N_{\text{Prot}}}  \sum_k^{N_j} \left (\sum_i^{N_{\text{Seq}}} P(S_i,\Gamma_j) \gamma^{i}_k - \gamma^{\text{Real}_{j}}_k\right)^2+ \nonumber\\
  E_{\text{Shannon}}\sum H(\epsilon)\log H(\epsilon) \label{Scoring_F}
\end{eqnarray}
where the index $i$ runs over the $N_{\text{Seq}}$ designed sequences for each protein $j$ of length $N_j$, the $E_{\text{Sol}}$ is the hydrophobicity scale of each residue (see Eq.S3 in the SM), while the $\gamma^i_k$'s are the contribution to the total energy of each residue calculated within the Caterpillar model. The last term instead guarantees that the Shannon entropy associated to the matrix elements $\epsilon_{kl}$ ($H$ are the histograms) is maximized to avoid an uniform matrix. We phenomenologically determined $E_{\text{Shannon}}=8.0$  for the scaling term to be a good value.  Note that here and in the following, energies are given in units of $k_{\rm B}T_{\rm {Ref}}$, where $T_{\rm {Ref}}$ is a reference temperature that sets the scale of the interactions, hence all simulation temperatures are given in units of $T_{\rm {Ref}}$. It is important to stress that $T_{\rm {Ref}}$ is not necessarily the folding temperature or the environment temperature, but all the energies can be rescaled to have $T_{\rm {Ref}}$ matching the physical temperature. In fact, in what follows we will show that all proteins studied fold approximately at the same temperature, one could think to rescale the energies to set the folding temperature to the one observed in nature. A schematic representation of the algorithm is reported in Fig.~\ref{MEP_Scheme}. 

To the best of our knowledge our work is the first of his kind to optimize the model parameters by reducing the differences between natural and designed sequences and, thanks to the MCP, is the simplest (in terms of the number of parameters needed) to successfully and quantitatively reproduce both sequences and structures of natural proteins to high precision. 

\begin{figure}[ht]
\includegraphics[width=1.00\columnwidth]{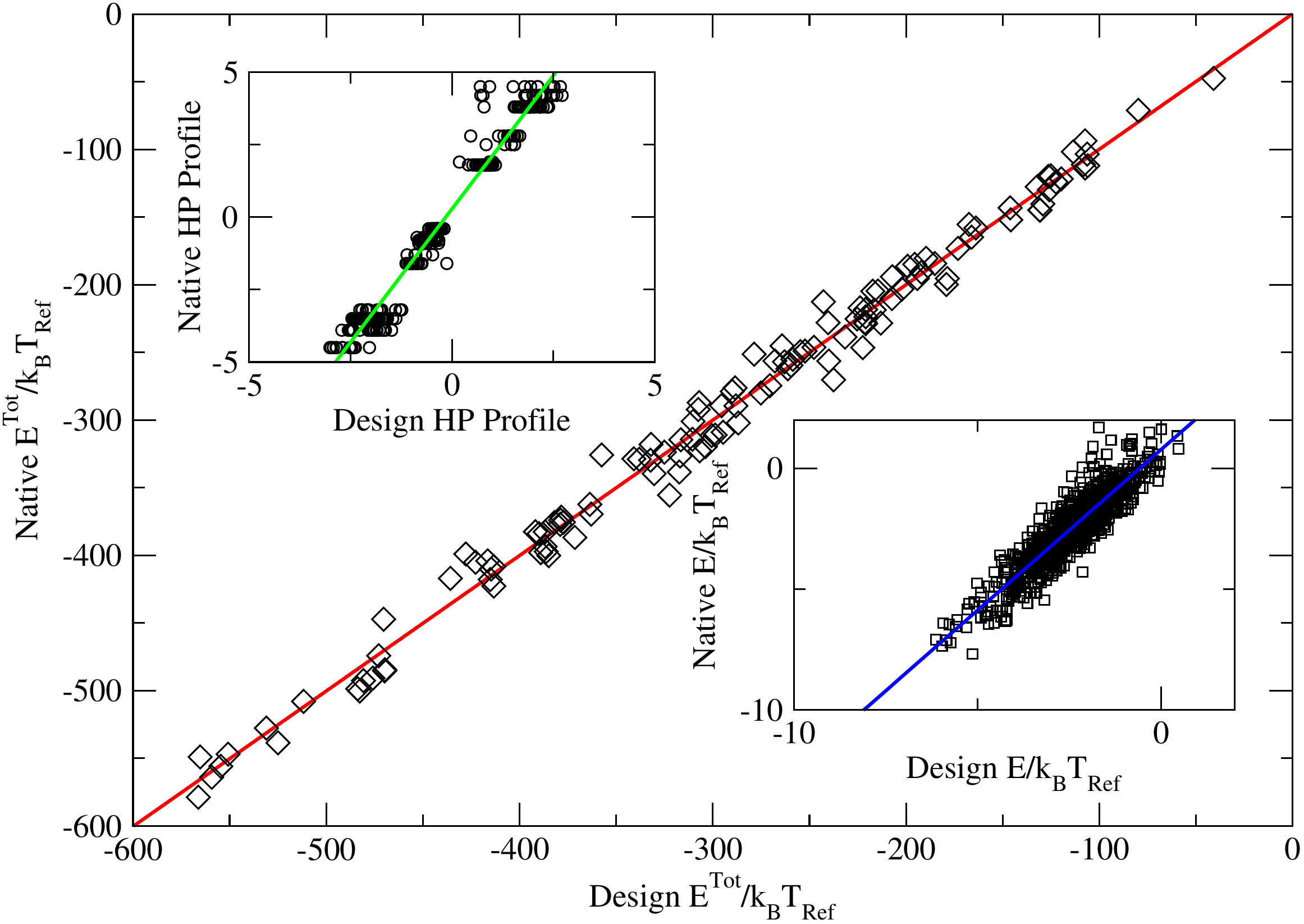}\caption{Comparison between the total residue energy $E^\text{Tot}/k_BT_{\text{Ref}}$ (Eq.S4 in SM) averaged over the designed sequences (abscissa) and the same energy calculated over the native sequences (ordinate). Each point corresponds to one protein in the data set and shows a strong linear trend verified by the fit (red line) with a correlation coefficient of $\sim 0.995$ and a slope of $\sim 1.000$ indicating that two energies are perfectly correlated. In the insets we show the comparison of the HP profiles (top left) and interaction energy $E/k_BT_{\text{Ref}}$ of each residue with all other (bottom right), this time each point corresponds to a single residue of each test protein. In both cases the data follow a remarkable linear trend (fits in green and blue lines respectively), and a positive correlation close to unity. For the HP profiles the correlation coefficient ($\sim 0.98$) indicates that when in natural proteins we find an hydrophobic residue also the design procedure will put one and vice versa. While the correlation coefficient ($\sim 0.90$) of $E/k_BT_{\text{Ref}}$ demonstrates that each natural residue has a very similar contribution to the total energy compared to the designed ones. A perfect match cannot be expected since natural sequences might have experience a selection pressure influenced by interactions not represented in the model, different environmental conditions or simply unknown functional requirements. Nevertheless the accordance is remarkable.}\label{Res_MEP}
\end{figure}

We began by selecting a protein training set from the Protein Data Bank (PDB)~\cite{Berman2000}, which includes all the proteins that obey the following conditions: a X-Ray structural resolution below 1.5~\AA, are made of single chains of length ranging from 20 to 200 residues, and do not contain any DNA or RNA. According to the stated conditions we selected 125 proteins (see Tab. S2 of the SM for the complete list of the PDB id's). It is important to stress that we did not select for specific experimental conditions, in particular pH and temperature during the measurements fluctuate significantly among the proteins in the set. In Fig.~\ref{Res_MEP} and Fig. S3 we plot the comparison of the natural to the designed sequences, the latter obtained with the MEP optimized interaction parameters. The plot shows strong correlation ($>0.9$) between the total energy of the designed (abscissa) and natural (ordinate) sequences, and between the profiles of the residue the HP profiles and the energy contribution (Fig.~\ref{Res_MEP} top and bottom insets and Fig. S3). \textit{Overall we can conclude that, for all 125 proteins in the training set, the designed proteins are very similar to natural proteins, demonstrating that our procedure can now be used to design realistic protein sequences.} We applied the MEP derived parameters to design 15 randomly selected from independent training sets~\cite{Kinch2011,Tsai2003}, and characterized by different secondary and tertiary motives. The top five resulting sequences for each target structures are listed in Tab. S4 of the SM. It is important to stress for this design we relaxed the constraint on the amino acid composition used during the optimization. Hence, the folding of the designed sequences does not depend on the previous knowledge of the natural amino acid composition, nevertheless the amino acids composition of the artificial sequences is similar to the natural one (see Tab. S3).  It has to be said that the artificial sequences appear unusual with repeats of the same amino acid (e.g. for 1gab \textit{WDDMIIRRRRFVVYYLWGSMTAEVEAEKGTNGFYYHHHDFGTKKKAQQQSNNL}). Such repeats  could be due to the approximations of the model, however it is important to remember that we did not include in the design any information about the function of the protein. In fact there is no reason to expect that natural sequences are the only one capable of folding, and we want to stress again that additional constraints applied during the design procedure would dramatically reduce the volume of the sequence space~\cite{Coluzza2012a} reducing the probability of repeats. We believe the latter to be the main cause of the repeats since as we will show below the model is capable of refolding also several natural sequences, demonstrating that in the model the presence of repeats is not necessary to stabilize natural protein structures. It is important to stress that during the last step of the MEP optimization the fewer sequences generated with fixed composition do not present the repeating patterns (see Table S5). However, this is an interesting problem and deserves a dedicated study that is beyond the objective of this work. Objective of ongoing research is also to experimentally test whether such sequences are capable of folding to the predicted target structures.

\begin{figure}[ht]
\includegraphics[width=1.00\columnwidth]{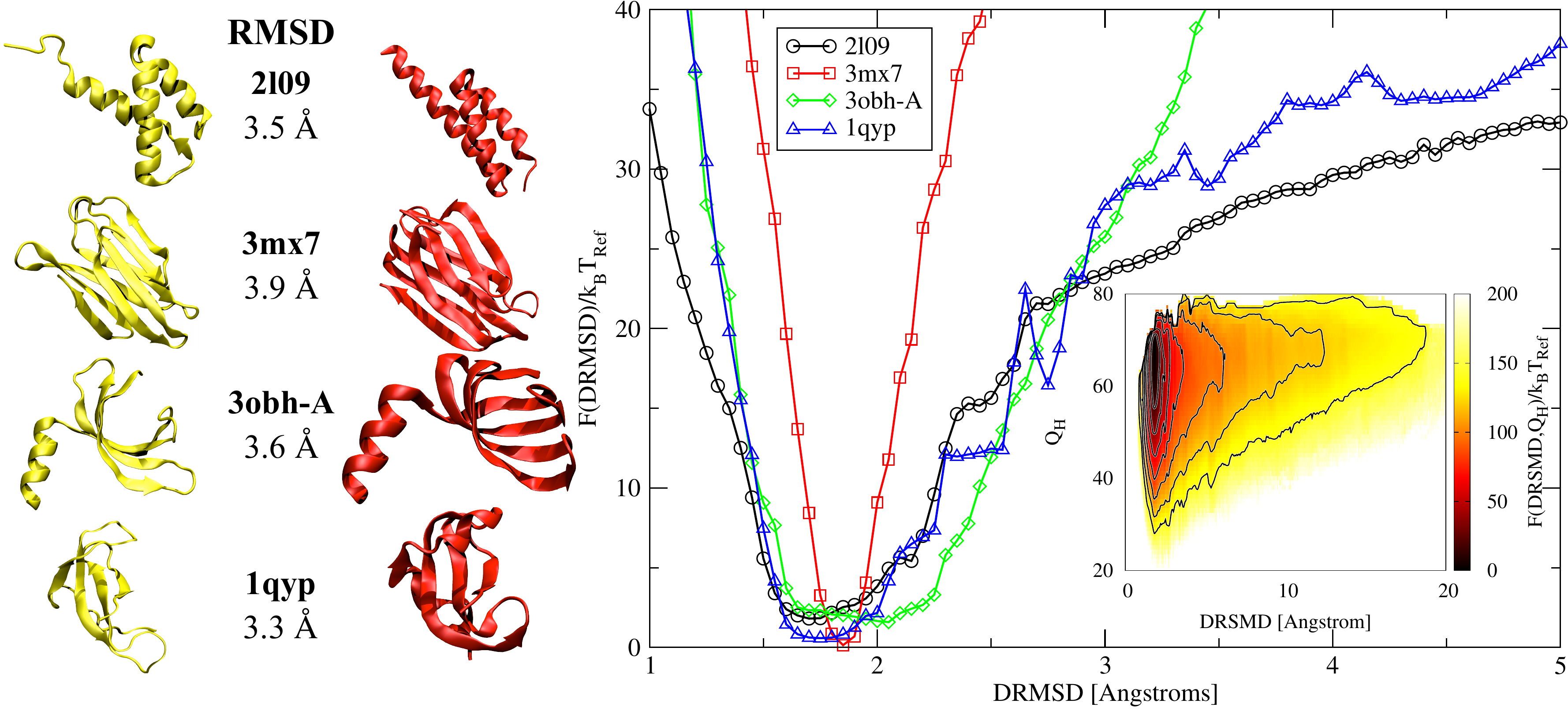}\caption{Folding free energy landscape $F(\text{DRMSD})/k_{\rm B}T_{\rm {Ref}}$ as a function of DRMSD of the four designed proteins (PDB ids 2l09, 3mx7, chain A of 3obh, and 1qyp). All profiles have a global minimum around $1.5$ and $2$~\AA~ DRMSD with a smooth funnelled shape. Due the approximations present in the model and to thermal fluctuations is  shifted with respect to $\text{DRMSD}=0$ (note that to the value $\text{DRMSD}=0$ of each profile will correspond a different native structure). Because of the definition of DRMSD, the smaller the value the fewer are the possible structures that can have this value of DRMSD. Ultimately, $\text{DRMSD}=0$ is possible only for the target structure itself. The funnelled profiles with single minimum implies that both an ensemble of arrested structures and a single alternative fold are less stable compared to the desired configuration. In the bottom right inset we plot the folding free energy landscape $F(\text{DRMSD},Q_{H})/k_{\rm B}T_{\rm {Ref}}$ for 3mx7 as a function of both the DRMSD and the number of hydrogen bonds $Q_H$, to give a visual example of the funnel nature of the folding landscapes. On the left we compare the experimentally determined structures (in yellow) with a typical folded  conformation selected as the sampled configurations with the lowest energy at the free energy minimum (in red). The RMSD value is indicated in the middle.}\label{Res_Design}
\end{figure}

In order to apply the model to the folding of both designed and natural sequences, we need to balance the residue energy term with the backbone hydrogen bond term ( parameter $\alpha$ in Eq.S4 in the SM). The energies can be rescaled by choosing the value of $\alpha$ for which designed sequences fold best to their target structures~\cite{Coluzza2011}. Hence, we selected four designed sequences from Tab. S4 (PDB ids 2l09,3mx7,chain A of 3obh, and 1qyp), and for each sequence we performed a refolding simulation (see SM) with different values of the rescaling parameter $\alpha$  in the range [0.05 to 1.0]. The best value of $\alpha=0.10\pm0.01 ~k_BT_{\text{Ref}}$ was the one for which all four proteins folded closer and smoother to the native state. In Fig.~\ref{Res_Design} we plot the refolding free energy $F(\text{DRMSD})/k_BT_{\text{Ref}}$ as a function of the distance root mean square displacement (DRMSD, see Appendix~DMRSD of the SM), obtained with the best energy value for $\alpha=0.10\pm0.01 ~k_BT_{\text{Ref}}$ for the four target proteins below the folding temperatures (estimated to be $T_F\approx 0.22$ for all proteins see SM for details). The plot shows for each protein a funnelled profile with a global minimum very close to the respective target structure ($\text{DRMSD} \in[1.5-2.0]$~\AA). So at least below the folding temperature the proteins seems to follow a downhill process. This observation would need a verification with a study of the folding dynamics. \textit{The refolding free energy profiles shown in Fig.~\ref{Res_Design} prove that realistic protein sequences with low frustration folding free energy landscapes can now be designed with a straightforward positive design scheme}.

We now have obtained the optimized parameters for our model: $\alpha=0.10\pm0.01~k_BT_{\text{Ref}}$, $\Omega=21.0\pm0.5$, $E_{\text{HOH}}=0.015\pm0.001$ and for the $\epsilon$ see Tab.S1 in the SM.

\begin{figure}[ht]
\subfigure{\includegraphics[width=0.49\columnwidth]{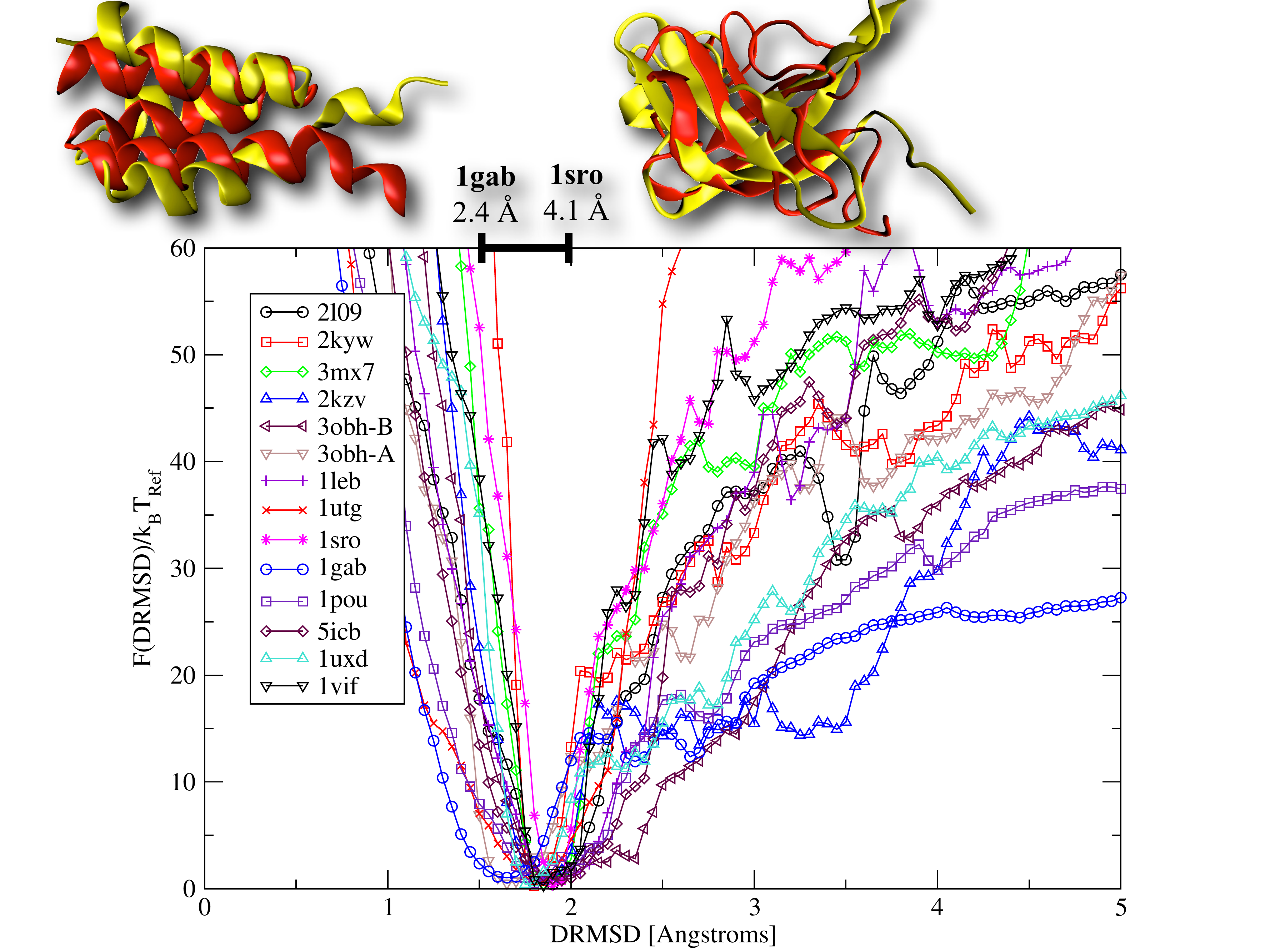}\label{Res_Fold}}
\subfigure{\includegraphics[width=0.49\columnwidth]{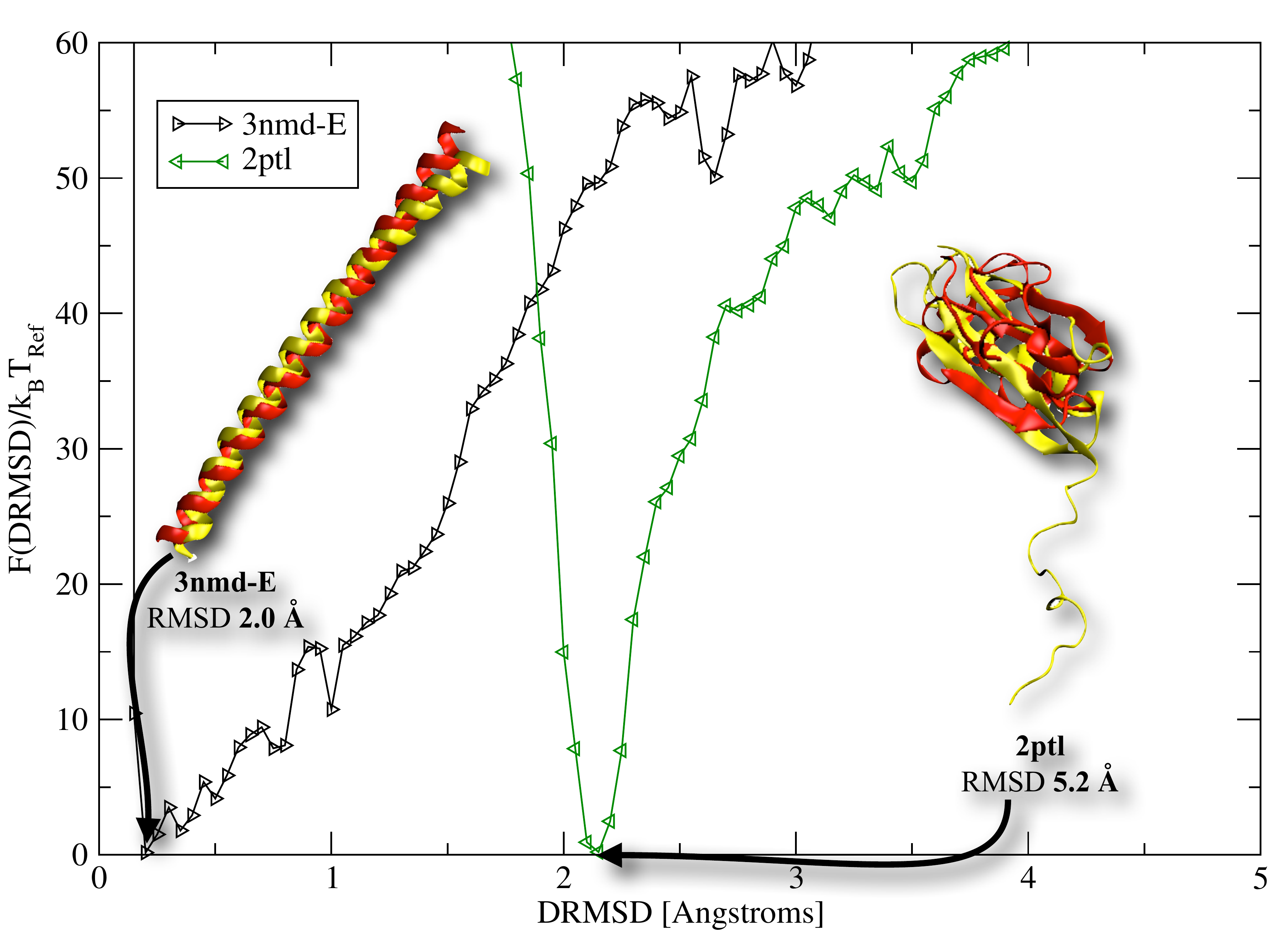}\label{Res_Fold_part}}\caption{Folding free energy landscape $F(\text{DRMSD})/k_{\rm B}T_{\rm {Ref}}$ of the 15 proteins set selected to test the accuracy of the MEP optimized parameters. The profiles have a common funnel shape and show a clustering of the free energy minima in the region $1.5$ and $2$~\AA~DRMSD  consistent with the results obtained for designed sequences. In b) we plot the free energies for proteins with the worst (2ptl) and the best (3nmd-E) distance of the folded structure from the native one. For the latter the free energy profile shows a minimum remarkably close to the native state probably due to the highly simplified structure of protein 3nmd-E.  The minimum of 2ptl, on the other hand, is located further away from the low DRMSD values than the other proteins. This apparent discrepancy is due to the definition of the DRMSD which includes the contribution from the $C_\alpha$ atoms located in the long unstructured tail from the residue 1 to 18. Since the probability of observing that particular conformation in solution is very low, it follows that the particular realization of the native structure has a large entropy penalty. However if we measure the overlap ignoring the contribution from the tail we see that the predicted structure  of the protein  core  is again reasonably close to the experimentally determined  one  ($\approx 5.2$~\AA~RMSD).  In the insets we compare the experimental structures (in yellow) super-imposed to the equilibrium configurations (in red), and we show that the proteins refolded with a precision between $2.4$ and $4.1$~\AA~RMSD.}
\end{figure}

The next logical step is to asses the behaviour of the model when refolding natural sequences and prove that folding as well can be performed to a quantitative level with the model. For this we randomly selected 15 proteins known to be difficult to fold  (from Tsai et al.~\cite{Tsai2003} and from the 9th edition of the well known Critical Assessment of Techniques for Protein Structure Prediction~\cite{Kinch2011}) and we performed folding simulations of their natural sequences. The results are plotted in Fig.~\ref{Res_Fold} and ~\ref{Res_Fold_part}, where we have superimposed all the computed free energy profiles. Although, the details of each profile might not be clearly visible, a first fundamental feature is apparent, namely the concentration of the free energy minima in the region between $1.5$ and $2$~\AA~DRMSD which remarkably is also the same regions observed for the design proteins. A second important result is the funnel shape common to all free energy profiles providing definite proof of the capability of the model of capturing the low frustration folding of natural proteins with a rather high precision. In fact when the predicted conformations of the folded states are compared to the experimentally determined structures, the two overlapped  with a precision between $2.4$ and $4.1$~\AA~RMSD (see top inset of Fig.~\ref{Res_Fold}) which is surprisingly accurate especially considering the simplicity of the model. It is important to note that the configurations with the lowest energy are not necessarily equal to the ones corresponding to the minimum of the free energy, however, in most cases, they are very similar. This is due to the strong directional nature of the hydrogen bonds which makes them very sensitive to thermal fluctuations. As a consequence, there are isolated structures that might have a lower energy but are not very stable at finite temperature.

To the best of our knowledge our coarse-grained protein model is the simplest, in terms of the number of parameters needed, with a transferable energy function capable of achieving such precision for the prediction of the native folded structures. Also it is one of the very few models that allows for both quantitative proteins design and folding, the latter demonstrated by free energy calculations. It is remarkable that low frustration sequences can be obtained with such a simple and universal design procedure, and that the folding of natural proteins shows funnelled free energy landscapes without the need of any potentials based on the native structure~\cite{GO1978}.

Although, the artificial sequences present some unnatural features like repetitions of some amino acids, the sequences designed with a natural amino acid composition share many features with the natural occurring ones, and the native structures of the latter are correctly predicted by our model. Hence, we expect that our designed proteins (see Tab. S4), once synthesized,  would fold to the structures used as design targets, which would also represent the ultimate and most important test of our methodology. We hope that our methodology will become an useful tool in experiments requiring  alterations of natural proteins, or the total redesign of target protein structures. Of course, constraints on the composition can always be applied to the design procedure with no major changes in the procedure. Moreover, the prediction power of the model gives us high confidence that our design methodology could be directly used to tackle important open problems of drug design, or used in a multi-scale approach where the results from our model could be refined with a more accurate but also a computationally more expensive protein model. 

Finally, this work not only extends our previous results obtained with the Caterpillar model, but also strengthens the connection among all our work on lattice heteropolymers and protein unrelated systems such as patchy polymers~\cite{Coluzza2012c,Coluzza2013}. The success of the same design strategy for all these systems demonstrates that the  minimum constraint principle is a sufficient condition to satisfy for the generalized design of low frustration sequences and the prediction of their proper native state. 

\begin{acknowledgments}
We would like to thank  Peter van Oostrum, Barbara Capone, Francisco Martinez-Veracoechea, Angelo Cacciuto and Christoph Dellago for fruitful discussions and a critical reading of the manuscript. All simulations presented in this paper were carried out on the Vienna Scientific Cluster (VSC).
\end{acknowledgments}

\section*{Supplementary Informations}

\setcounter{equation}{0}
\setcounter{figure}{0}
\setcounter{page}{1}
\setcounter{table}{0}
\renewcommand{\theequation}{S\arabic{equation}}
\renewcommand{\thefigure}{S\arabic{figure}}
\renewcommand{\thepage}{S\arabic{page}}
\renewcommand{\thetable}{S\arabic{table}}

\subsection{Details of the Caterpillar interaction potential}\label{ModelDetails}


The Caterpillar model ~\cite{Coluzza2011} is a 5-atom model of the protein backbone (see Fig.~1).  The degrees of freedom of the
model are the torsional angles $\phi_1$ and $\phi_2$; all other structural parameters are kept fixed at values from the literature~\cite{Creighton1992}.  
Backbone hydrogen bonds are modeled with a 10-12 Lennard-Jones type potential using the expression~\cite{Irback2000}
\begin{equation}
E_{H}=-\epsilon_{H} \left(\cos{\theta_1}\cos{\theta_2}\right)^\nu\left[5\left(\frac{\sigma}{r_{\textrm{OH}}}\right)^{12}-6\left(\frac{\sigma}{r_{\textrm{OH}}}\right)^{10}\right],
\label{H-Bond-Int}
\end{equation}
\noindent where $r_{\textrm{OH}}$ is the distance between the hydrogen atom of the amide group (NH) and the oxygen atom of the carboxyl group (CO) of the main chain. We set $\sigma=2.0\:\text{\AA}$, $\epsilon_{H}=-3.1~k_BT_\text{Ref}$, and $\nu=2$~\cite{Irback2000}. 
The side chain interactions are represented by and effective  $C_{\alpha}$-$C_{\alpha}$ sphere-sphere interaction energy given by 
\begin{equation} 
E_{ij}\left(r_{ij}\right)=\epsilon_{ij}\Gamma(r_{ij})=\epsilon_{ij}\frac{1}{1 + \textrm{e}^{-\left(r_{\textrm{max}} - r_{ij} \right)/W}}
\label{CalphaInt}
\end{equation}
\noindent where $W=0.4 \text{\AA}$, $r_{ij}$ is the distance between the $C_{\alpha}$ atoms at the centres of spheres $i$ and $j$ and $r_{\textrm{max}}$ ($r_{\textrm{max}}=12\:\text{\AA}$) is the distance at which $E_{ij}=\epsilon_{ij}/2$. The $\epsilon_{ij}$ are the elements a 20 by 20 matrix each defining the strength of the interaction of each type of amino acid with the others (see Tab.~\ref{Matrices} for the values optimized with the MEP). 

The residue-solvent interaction is modelled as a simple energy penalty towards surfaces exposure of hydrophobic amino acids; the expression has the form
\begin{equation}
\begin{array}{lll}
 E_{\text{Sol}}(\Omega-\Omega_i) &= &\left\lbrace
\begin{array}{lll}
 \epsilon^i_{\text{Sol}}\left[\Omega-{\Omega_i}\right] & \Omega_i \lesseqgtr \Omega &  \epsilon^i_{\text{Sol}}\gtrless0\\
0 & \Omega_i\gtrless\Omega &  \epsilon^i_{\text{Sol}} \gtrless0\\

\end{array}\right.\\
&&\Omega_i =\sum_j \Gamma(r_{ij})
\end{array},\label{equation3}
\end{equation}
where $\Gamma(r_{ij})$ is given in Eq.~\eqref{CalphaInt}, $\Omega$ is the threshold for the number of contacts in the native structure above which the amino acid is considered to be fully buried and the $\epsilon^i_{\text{Sol}}$ are taken from the  the Dolittle hydrophobicity index~\cite{Dolittle1989} and are positive for hydrophobic amino acids and negative of the hydrophilic ones. The interaction penalises the exposure (burying) of hydrophobic (hydrophilic) residues above $\Omega$.
The formation of sulphur bridges as well as Proline rigid bonds is not included. The total energy of a protein $E$ is then given by:

\begin{eqnarray}
E&=&E_{H}+\alpha E(\epsilon,E_{\text{HOH}},\Omega) \nonumber \\
 &&E(\epsilon,E_{\text{HOH}},\Omega)=\sum^{N}_{kl} \epsilon_{kl}\Gamma(r_{kl})+E_{\text{HOH}}\sum^{N}_{k}E_{\text{Sol}}(\Omega-\Omega_k)\label{total_hamiltonian}
\end{eqnarray}
where $E_{H}$ in Eq.~\ref{H-Bond-Int} is the total energy of the backbone hydrogen bonds, the $\alpha$ scaling factor is necessary to balance the two contributions to the total energy. Large values of $\alpha$  will tend to break the minimum constraint principle, while small values will over-favour the hydrogen bond term inducing the formation of just helices. Finally  $E_{\text{HOH}}$, like $\alpha$, rescales the Dolittle hydrophobicity~\cite{Dolittle1989} scale appropriately.

\subsection*{Max Entropy Derivation} \label{MaxEntrDeriv}

We will consider a set of $j\in{1,\dots,N_{\text{Prot}}}$ proteins each of length $N_j$. To each protein an ensemble of  $i\in{1,\dots,N_{\text{Seq}}}$ sequences is designed. Hence the probability $P(S_i,\Gamma_j)$ of having a sequence $i$ on structure $j$ is given by the Boltzmann weight:
\begin{equation}
 P(S_i,\Gamma_j)=\frac{\exp^{-\beta E(\epsilon,E_{\text{HOH}},\Omega)[S_i,\Gamma_j]}}{\sum_i^{N_{\text{Seq}}} \exp^{-\beta E(\epsilon,E_{\text{HOH}},\Omega)[S_i,\Gamma_j]}},
\end{equation}

\noindent where $E(\epsilon,E_{\text{HOH}},\Omega)[S_i,\Gamma_j]$  is the residue interactions energy function in Eq.~\ref{total_hamiltonian} calculated for sequence $S_i$ and conformation $\Gamma_j$. It is important to stress that since during the design procedure the backbone degrees of freedom of the target structure are kept frozen, the  backbone hydrogen bonds are also frozen and so do not play a role in the optimization procedure. 

The objective is to determine the unknown parameters $\epsilon$, $E_{\text{HOH}}$, and $\Omega$ following the MEP procedure. Hence, we need to maximize the entropy associated to the probability distribution $P(S_i,\Gamma_j)$, under the constraint that the designed sequences are as close as possible to the natural ones. Our choice for the constraints is based on the results presented in our previous publication~\cite{Coluzza2011}, were we only optimized $E_{\text{HOH}}$, and $\Omega$  to  match the HP profiles among designed and natural sequences for just protein 1ctf. No optimization of the interaction matrix $\epsilon$ was performed. Nevertheless, the matching was successful, and the novel energy function was capable of refolding the natural sequence of 1ctf. 
Let us start by introducing the entropy $S$ associated to $P(S_i,\Gamma_j)$
\begin{equation}
 S=-\sum_j^{N_{\text{Prot}}} \sum_i^{N_{\text{Seq}}} P(S_i,\Gamma_j) \ln P(S_i,\Gamma_j)
\end{equation}
For simplicity we show the derivation of the maximum entropy principle for  $P(S_i,\Gamma_j)$ only under the constraints that  designed proteins have an average HP profile close to the natural one, and that $P(S_i,\Gamma_j)$ is normalized.
In order to do so we are going to make use of the Lagrange multiplier method, which means that the maximization process is equivalent to find the extremal of the function $\Lambda$ defined as follows:
\begin{equation}
 \Lambda=S+\sum_j^{N_{\text{Prot}}}  \sum_k^{L} \lambda_{jk} \left(\sum_i^{N_{\text{Seq}}} P(S_i,\Gamma_j) \alpha^{i}_k - \alpha^{\text{Real}_{j}}_k\right)+\sum_j^{N_{\text{Prot}}} \gamma_j \left(Z_j-1\right)
\end{equation}
where the $\lambda_{jk}$ and $\gamma_j$ are the Lagrange multiplier and $Z_j=\sum_i^{N_{\text{Seq}}} P(S_i,\Gamma_j)$ is the partition function.
We now have to derive the $\Lambda$ function with respect to $P$ keeping the Lagrange multiplier constant and look for the maximum.
\begin{equation}
 \frac{d\Lambda}{d P(S_i,\Gamma_j)}=-\ln P(S_i,\Gamma_j) +1 + \sum_k^{L} \lambda_{jk}  P(S_i,\Gamma_j) \alpha^{i}_k + \gamma_j = 0
\end{equation}
which gives for the $P(S_i,\Gamma_j)$ the following expression
\begin{eqnarray}
  P(S_i,\Gamma_j) &=& e^{\gamma_j+1}e^{\sum_k^{L} \lambda_{jk}\alpha^{i}_k}\\
\sum_i^{N_{\text{Seq}}} P(S_i,\Gamma_j)&=&1\\
e^{\gamma_j+1}&=&\frac{1}{\sum_i^{N_{\text{Seq}}} e^{\sum_k^{L} \lambda_{jk}\alpha^{i}_k}}\\
P(S_i,\Gamma_j) &=& \frac{e^{\sum_k^{L} \lambda_{jk}\alpha^{i}_k}}{\sum_i^{N_{\text{Seq}}} e^{\sum_k^{L} \lambda_{jk}\alpha^{i}_k}}\nonumber\\
&=&\frac{e^{\sum_k^{L} \lambda_{jk}\alpha^{i}_k}}{Z'_j}
\end{eqnarray}
Now that we have the relation that connects the Lagrange multiplier with the probability distribution $P$ we can express the function $\Lambda$  in terms of the Lagrange multipliers only
\begin{eqnarray}
 \Lambda&=&-\sum_j^{N_{\text{Prot}}} \sum_i^{N_{\text{Seq}}}  \frac{e^{\sum_k^{L} \lambda_{jk}\alpha^{i}_k}}{Z'_j} \left( \sum_k^{L} \lambda_{jk}\alpha^{i}_k-\ln Z'_j \right)\nonumber\\
&&+\sum_j^{N_{\text{Prot}}}  \sum_k^{L} \lambda_{jk} \left(\sum_i^{N_{\text{Seq}}} \frac{e^{\sum_k^{L} \lambda_{jk}\alpha^{i}_k}}{Z'_j}\alpha^{i}_k - \alpha^{\text{Real}_{j}}_k\right)
\end{eqnarray}

If we now rearrange the terms inside the sums

\begin{eqnarray}
 \Lambda&=&-\sum_j^{N_{\text{Prot}}} \sum_i^{N_{\text{Seq}}}  \frac{e^{\sum_k^{L} \lambda_{jk}\alpha^{i}_k}}{Z'_j}  \sum_k^{L} \lambda_{jk}\alpha^{i}_k+\sum_j^{N_{\text{Prot}}}\ln Z'_j \nonumber\\
&&+\sum_j^{N_{\text{Prot}}}  \sum_k^{L} \lambda_{jk} \sum_i^{N_{\text{Seq}}} \frac{e^{\sum_k^{L} \lambda_{jk}\alpha^{i}_k}}{Z'_j}\alpha^{i}_k -\sum_j^{N_{\text{Prot}}}  \sum_k^{L} \lambda_{jk} \alpha^{\text{Real}_{j}}_k\nonumber\\
&=&-\sum_j^{N_{\text{Prot}}} \sum_k^{L} \lambda_{jk} \sum_i^{N_{\text{Seq}}}  \frac{e^{\sum_k^{L} \lambda_{jk}\alpha^{i}_k}}{Z'_j}  \alpha^{i}_k+\sum_j^{N_{\text{Prot}}}\ln Z'_j \nonumber\\
&&+\sum_j^{N_{\text{Prot}}}  \sum_k^{L} \lambda_{jk} \sum_i^{N_{\text{Seq}}} \frac{e^{\sum_k^{L} \lambda_{jk}\alpha^{i}_k}}{Z'_j}\alpha^{i}_k -\sum_j^{N_{\text{Prot}}}  \sum_k^{L} \lambda_{jk} \alpha^{\text{Real}_{j}}_k\nonumber\\
&=&\sum_j^{N_{\text{Prot}}}\ln Z'_j-\sum_j^{N_{\text{Prot}}}  \sum_k^{L} \lambda_{jk} \alpha^{\text{Real}_{j}}_k
\end{eqnarray}

We can now derive the condition for the Lagrange multiplier to maximize the functional $\Lambda$
\begin{eqnarray}
 \frac{\partial \Lambda}{\partial \lambda_{jk}}&=&\frac{1}{Z'_j}\frac{\partial Z'_j}{\partial \lambda_{jk}}-\alpha^{\text{Real}_{j}}_k\nonumber\\
&=& \frac{1}{Z'_j} \sum_i^{N_{\text{Seq}}} \alpha^{i}_k e^{\sum_k^{L} \lambda_{jk}\alpha^{i}_k}-\alpha^{\text{Real}_{j}}_k=0 \label{Solution_MEP}
\end{eqnarray}
This result can be interpreted as follows: the distribution generated by the Lagrange multiplier that makes the average hydrophobic profile equal to the natural one, is also the one that maximizes the entropy. Hence, we selected the following scoring function:
\begin{eqnarray}
  F_{\text{score}}=\sum_j^{N_{\text{Prot}}}  \sum_k^{N_j} \left (\sum_i^{N_{\text{Seq}}} P(S_i,\Gamma_j) E_{\text{Sol}}^{ik} - E_{\text{Sol}}^{\text{Real}_{jk}}\right)^2 + \nonumber\\ \sum_j^{N_{\text{Prot}}}  \sum_k^{N_j} \left (\sum_i^{N_{\text{Seq}}} P(S_i,\Gamma_j) \gamma^{i}_k - \gamma^{\text{Real}_{j}}_k\right)^2+ \nonumber\\
  E_{\text{Shannon}}\sum H(\epsilon)\log H(\epsilon) 
\end{eqnarray}
where $E_{\text{Sol}}$ is the hydrophobicity scale per residue (see Eq.~\ref{equation3}), $\gamma^i_k=\sum^{N_j}_{k} \epsilon_{kl}\Gamma(r_{kl})+E_{\text{HOH}}\sum^{N}_{k}E_{\text{Sol}}$'s are the contribution to the total energy per residue, and  $E_{\text{Shannon}}=8.0$. Hence, the scoring function is just a  comparison of the  designed  hydrophobic and energy profiles averaged over all designed sequences to the real profile. Since protein design can be performed in parallel, each averaging step is very fast. Nevertheless, before reaching total convergence of the parameters we had to perform $\sim 10^8$ iterations for a total of three weeks of computations on 250 AMD Opteron 6132 HE, 2.2 GHz each with 8 cores . It is important to note that, in order to speed up the sampling of such systems, inevitably rich in local energy traps, we used the Virtual Move Parallel Tempering scheme~\cite{Coluzza2005}.

\subsection*{Free Energy Calculations}

\begin{figure}[ht]
\includegraphics[width=1.00\columnwidth]{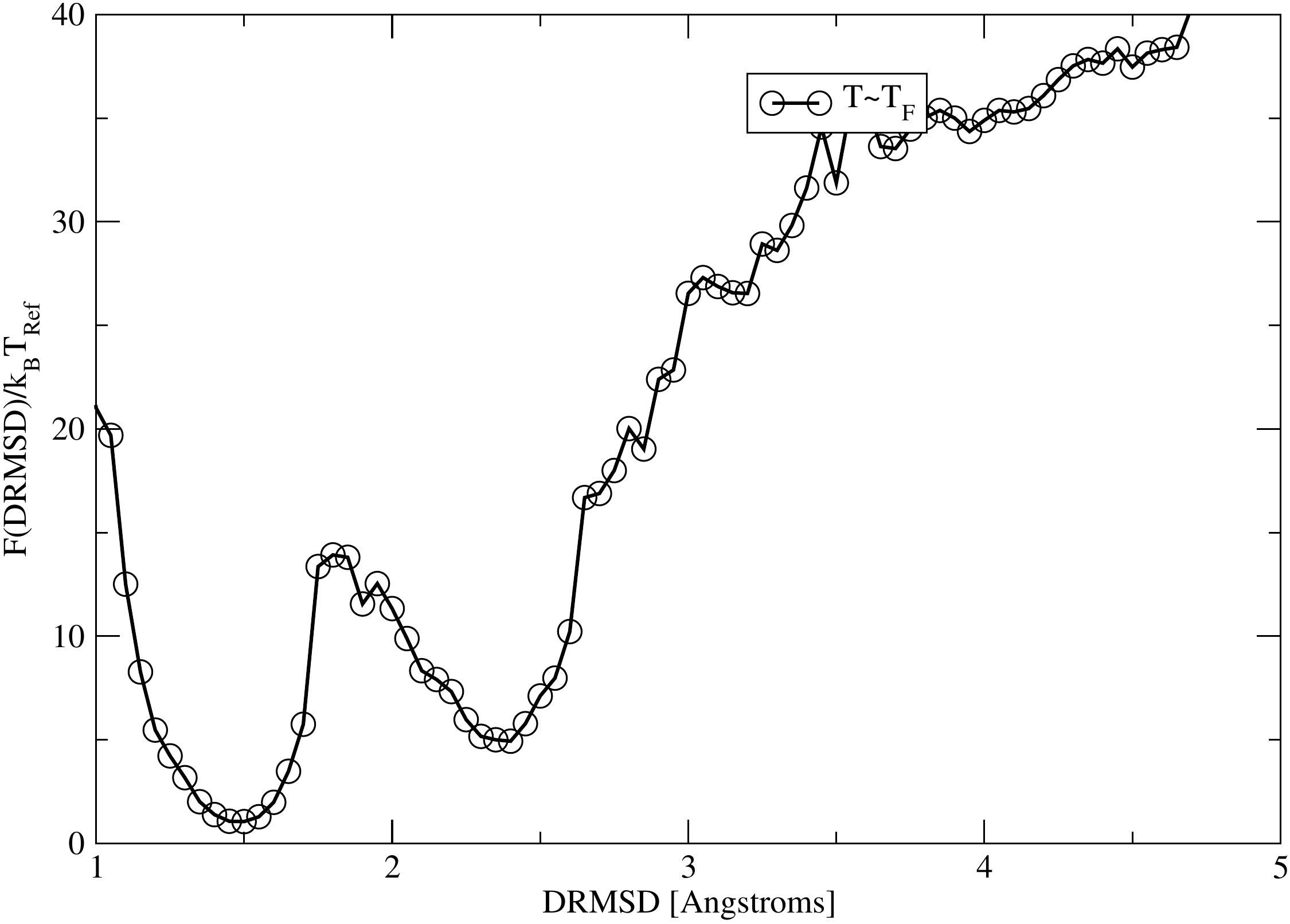}\caption{Folding free energy landscape $F(\text{DRMSD})/k_{\rm B}T_{\rm {Ref}}$ as a function of DRMSD of the designed protein PDB ids 1CTF close to the folding temperature}\label{1CTF}
\end{figure}

In order to test the folding behaviour of the designed and natural sequences we used two algorithms that we refer to as DESIGN and FOLDING to design new sequences and to compute the refolding free energies of both designed and natural sequences. Both DESIGN and FOLDING are described in our previous work~\cite{Coluzza2011}. The FOLDING simulation starts from a fully stretched chain and the configurational space is explored by means of pivot moves around the dihedral angles and by a crankshaft move~\cite{Frenkel2002b}. The latter consists in a rotation of all the atoms between two randomly chosen $C_\alpha$ carbons. The rotation is then performed around the axis connecting the two $C_\alpha$ atoms. When this move is performed on two consecutive $C_\alpha$'s it rotates the rigid body composed by the atoms C,N, H and O, and helps to equilibrate locally the hydrogen bonds. It is important to notice that the crankshaft move will distort the $\widehat{C C_\alpha N}$ angle which is kept close to his equilibrium value by a strong ($50~k_BT/\text{\AA}^2 $ elastic constant) spring, hence moves that case large distortion of the angle are rejected.
During a FOLDING simulation the sampled configurations are grouped in ensembles defined by the free energy $F\left({\rm DRMSD}\right)$ as a function of DRMSD to the target structure. For a well-designed sequence and a natural folding sequence will fold into structures fluctuating around the target structure corresponding to ${\rm DRMSD}=0$. In order to increase the sampling of the free energy for the FOLDING simulations we will use the Virtual Move Parallel Tempering (VMPT) scheme~\cite{Coluzza2005} (see SI) for the range of temperatures $[0.40, 0.31, 0.22, 0.20, 0.18, 0.16, 0.13, 0.12, 0.11, 0.10, 0.09, 0.08, 0.07, 0.06, 0.05, 0.04]$. We could not determine precisely the folding temperatures $T_F$  for all the proteins studied  and we only estimated it to be between the highest temperature where the protein is still folded and the one above where the protein is unfolded.  However and exact estimate is not essential for the scope of this work, but we plan in the future to study the folding transition more in detail. What we observed in our previous publication~\cite{Coluzza2011} for the protein 1CTF is that close to $T_F$ folded state has approximately the same free energy as the high temperature disordered globular state~\ref{1CTF}. It is important to stress that the Virtual Move Parallel Tempering biasing scheme is designed to allow the system to explore as much as possible of the free energy landscape and not to drive the system towards the minimum, which means that even if we would start form the folded configuration, the simulation would quickly evolve away from the global minimum.

\subsection*{DRMSD \label{sec:drmsd}} 

\begin{figure}[ht]
\subfigure[]{\includegraphics[width=.49\columnwidth]{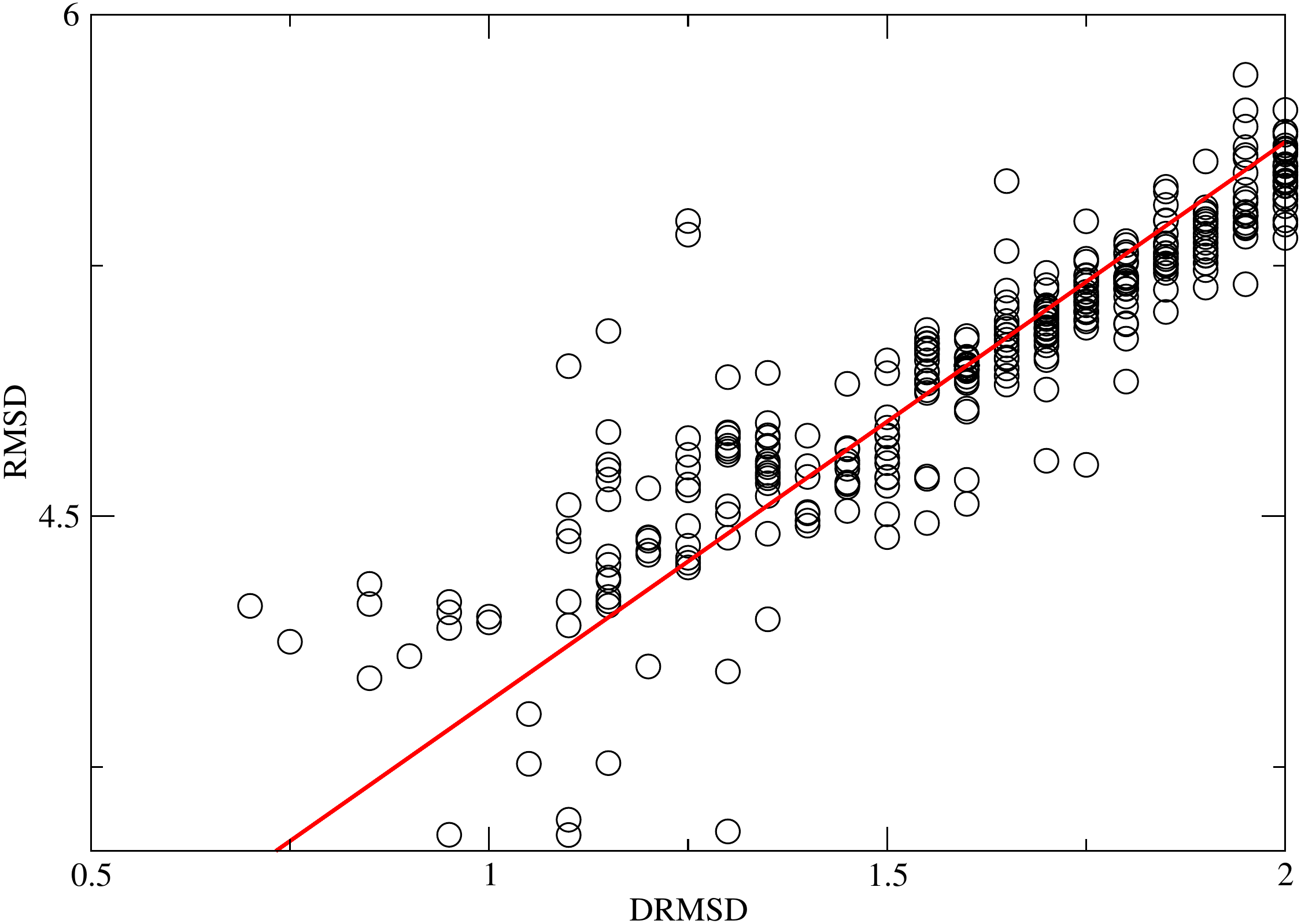}\label{Corr_DRMSD_vs_RMSD}}
\subfigure[]{\includegraphics[width=.49\columnwidth]{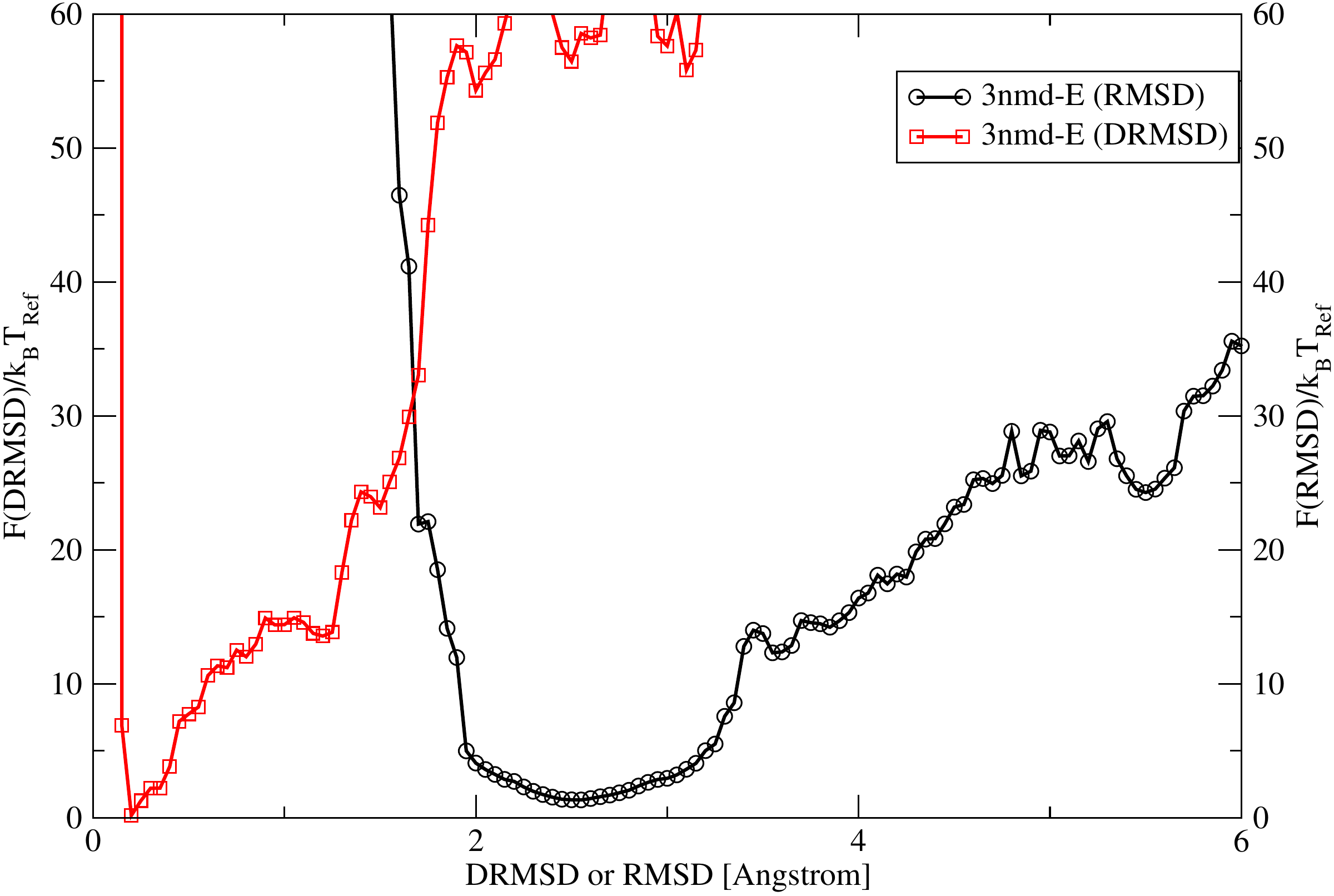}\label{DRMSD_vs_RMSD}}
\caption{On the left: correlation plot between the DRMSD and the RMSD collective variable. The estimated correlation coefficient  from a linear regression fitting (in red) is $\approx0.8$ which increases to $\approx0.98$, if we exclude the configurations for values of DRMSD$<1.5 \AA$ which is below the model resolution, indicating that the free energy profile should be qualitatively similar if the states are projected over RMSD instead of DRMSD. On the right: Free Energy folding profile of the protein 3NMD-E projected over the collective variables DRMSD and RMSD. The profiles are not identical because the RMSD is more sensitive to local distortions of the protein with respect to the DRMSD. This is also demonstrated by the wider free energy minimum which reflects the thermal fluctuations. However, overall the qualitative shape of the profiles is very similar with between each other in particular since both have a clear global free energy minimum.}
\end{figure}

The distance root mean square displacement is a standard collective variable used in the field of protein folding to measure the state of the folding transition. When a target structure is given the DRMSD is defined as 
\begin{equation}
\text{DRMSD}=\frac{1}{N}\sqrt{\sum_{ij}\left(\left|\Delta \vec{r}_{ij}\right| - \left|\Delta \vec{r}^{\;T}_{ij}\right| \right)^2}, \label{DRMSD}
\end{equation}
where $\vec{r}_{ij}$ is the distance between the sphere $i$ and $j$ while $\vec{r}^{T}_{ij}$ is the same distance calculated over the target structure, and $N$ is the chain length. According to Eq.\eqref{DRMSD}, DRMSD$=0$ is possible only when the chain and the target structures are identical. Any structural difference will correspond to larger values of DRMSD, and the larger the value of DRMSD the larger is the number of structures that share the same DRMSD from the target. In order to justify the use of DRMSD instead of more commonly use of RMSD we measured the correlation between the two collective variables. In Fig~\ref{DRMSD_vs_RMSD} we compare the two quantities and we measured the correlation faction near the free energy minimum. The results indicate that the two quantities are highly correlated for DRMSD$>1.5 \AA$ which is compatible with the current resolution of the caterpillar model.

\bibliography{library}
\pagebreak

\begin{figure}[ht]
\includegraphics[width=1.00\columnwidth]{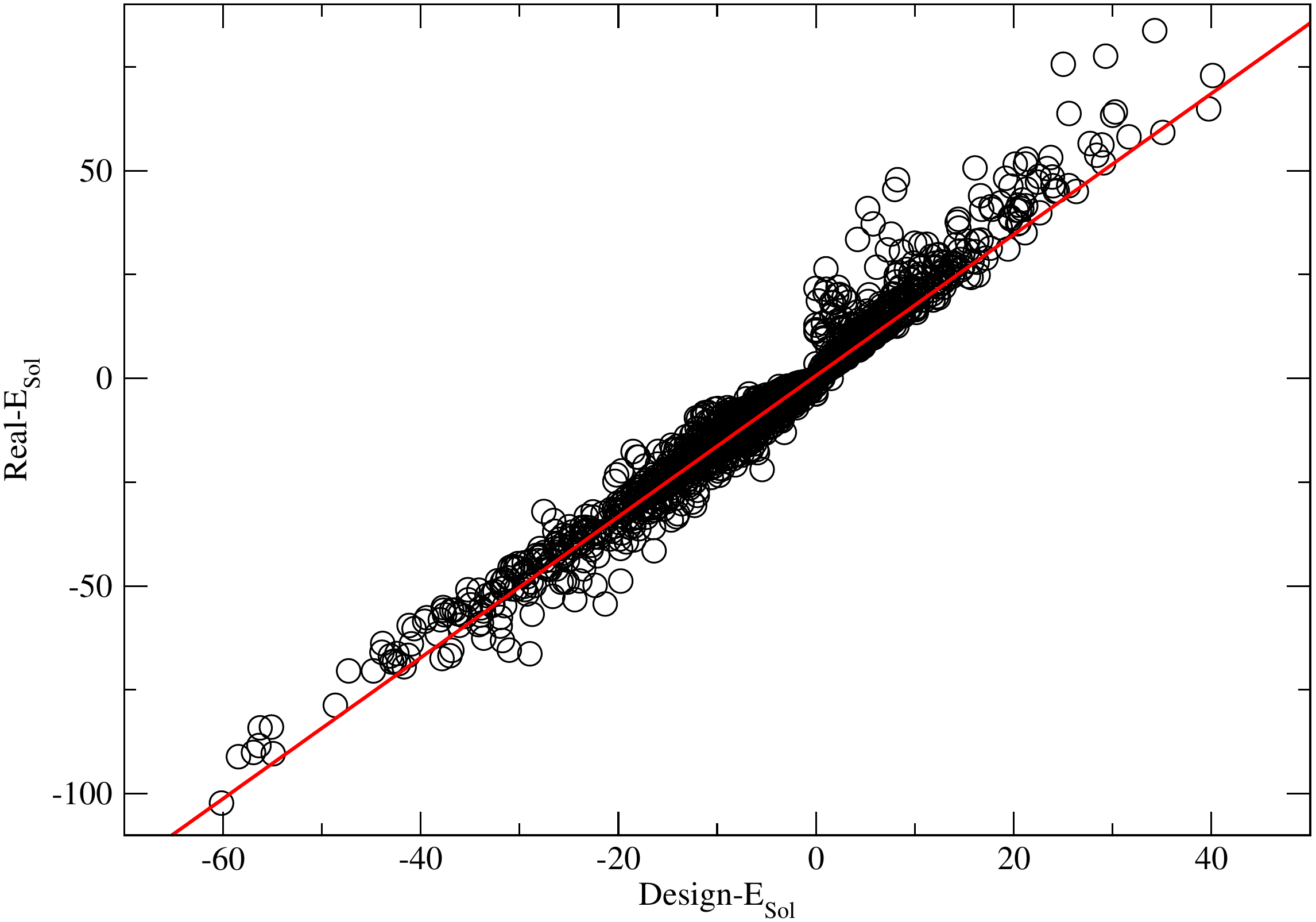}\caption{Correaltion between designed  and real $E_{sol}$ profiles. The correlation coefficient has been estimated from a linear regression fitting (in red) to be be very high $\approx 0.98$}\label{ESOL_profiles}
\end{figure}

\pagebreak
\begin{sidewaystable}
\caption{Optimized values of the residue-solvent $ \epsilon_{\text{Sol}}$ and residue-residue $\epsilon(S_k)(S_l)$ interaction parameters. The uncertainty on the values is $\approx \pm 0.01$}
\begin{center}
\resizebox{1.0\textheight}{!}{
\begin{tabular}{|c|c|c|c|c|c|c|c|c|c|c|c|c|c|c|c|c|c|c|c|c|c|}
\hline
Restype & $ \epsilon_{\text{Sol}}$ &\multicolumn{20}{c|}{$\epsilon(S_k)(S_l)$}\\
\hline
ALA &   1.80 &  0.10 & \multicolumn{19}{c|}{}\\
CYS &   2.50 & -0.10 &  0.39  & \multicolumn{18}{c|}{}\\
ASP &  -3.50 & -0.70 & -0.09  & -0.23  & \multicolumn{17}{c|}{}\\
GLU &  -3.50 & -0.23 & -0.44  & -0.17  & -0.32  & \multicolumn{16}{c|}{}\\
PHE &   2.80 & -0.22 & -0.92  &  0.41  &  0.32  & -0.33  &  \multicolumn{15}{c|}{}\\
GLY &  -0.40 & -0.45 & -0.18  &  0.51  & -0.00  & -0.51  & -0.37  & \multicolumn{14}{c|}{}\\
HIS &  -3.20 &  0.17 &  0.89  & -0.89  &  0.87  & -0.82  & -0.05  &  0.63  & \multicolumn{13}{c|}{}\\
ILE &   4.50 &  0.29 &  0.06  & -0.02  & -0.51  & -0.59  &  0.08  & -0.77  & -0.38  & \multicolumn{12}{c|}{}\\
LYS &  -3.90 & -0.02 &  0.63  &  0.13  & -0.47  & -0.85  & -0.99  & -0.52  &  0.23  &  0.19  &  \multicolumn{11}{c|}{}\\
LEU &   3.80 & -0.50 & -0.11  & -0.27  & -0.55  & -0.22  & -0.33  & -0.36  & -0.40  &  0.60  &  0.24  & \multicolumn{10}{c|}{}\\
MET &   1.90 &  0.46 &  0.67  &  0.59  & -0.92  & -0.29  & -0.48  & -0.26  &  0.02  &  0.69  &  0.01  &  0.89  & \multicolumn{9}{c|}{}\\
ASN &  -3.50 & -0.76 &  1.04  & -0.15  & -0.93  &  0.14  &  0.20  & -0.34  & -0.37  & -0.90  &  0.08  &  0.53  &  0.33 & \multicolumn{8}{c|}{}\\
PRO &  -1.60 & -0.72 &  0.49  &  0.03  & -0.46  & -0.64  &  0.82  & -0.19  &  0.05  &  0.80  & -0.70  & -0.41  & -0.48 & -0.57 &  \multicolumn{7}{c|}{}\\
GLN &  -3.50 & -0.83 &  0.09  & -0.29  & -0.64  & -0.35  & -0.69  &  0.17  &  0.30  & -0.87  & -0.80  &  0.68  & -0.66 &  0.74 &  0.34 &  \multicolumn{6}{c|}{}\\
ARG &  -4.50 &  0.36 &  0.40  & -0.73  &  0.56  & -0.80  & -0.60  & -0.94  & -0.38  & -0.12  & -0.88  & -0.49  & -0.85 & -0.79 & -0.20 &  0.12 &  \multicolumn{5}{c|}{}\\
SER &  -0.80 & -0.62 & -0.76  & -0.39  & -0.14  & -0.18  &  0.03  & -0.78  & -0.20  & -0.85  & -0.54  &  0.65  & -0.01 & -0.39 &  0.16 &  0.35 &  0.27 &  \multicolumn{4}{c|}{}\\
THR &  -0.70 & -0.12 & -0.99  & -0.24  & -0.44  & -0.52  & -0.63  & -0.74  &  0.57  & -0.07  & -0.49  & -0.07  &  0.91 &  0.53 & -0.78 &  0.26 &  0.15 &  0.54 &  \multicolumn{3}{c|}{}\\
VAL &   4.20 &  0.42 & -0.63  & -0.47  &  0.09  &  0.37  & -0.97  &  0.11  & -0.06  & -0.56  & -0.35  & -0.16  &  0.25 & -0.81 & -0.14 & -0.53 & -0.96 & -0.58 & -0.03 &  \multicolumn{2}{c|}{}\\
TRP &  -0.90 & -0.60 & -0.02  &  0.23  & -0.29  &  0.48  &  0.10  &  0.50  & -0.42  & -0.47  & -0.31  & -0.68  &  0.51 & -0.09 &  1.07 & -0.55 &  0.93 & -0.22 & -0.72 & 0.28 & \\
TYR &  -1.30 & -0.57 &  0.85  &  0.06  & -0.74  & -0.10  & -0.99  & -0.26  & -0.66  & -0.98  &  0.87  & -0.60  & -0.32 &  0.96 &  0.97 & -0.96 &  0.43 & -0.89 &  0.30 & 0.21 & 0.33 \\
\hline
\multicolumn{2}{|c|}{} & ALA & CYS & ASP & GLU & PHE & GLY & HIS & ILE & LYS & LEU & MET & ASN & PRO & GLN & ARG & SER & THR & VAL & TRP & TYR \\

\hline

\end{tabular}
}
\label{Matrices}
\end{center}
\end{sidewaystable}

\pagebreak

\begin{table}[ht]
\caption{List of PDB id's used as training set for the maximum entropy parameters optimization }
\begin{center}
\begin{tabular}{ccccc}
1WVH & 1MSI & 1LU4 & 1IFC &3EAZ \\
3B79 & 3EUR & 1TUA & 1ZCE &3JVE \\
1ZHV & 2VQ4 & 2BVV & 1T3Y &1TP6 \\
3Q6L & 3CX2 & 3F2Z & 3HP4 &2GRC \\
2E3H & 2GS5 & 1N7E & 2ESO &2WJ5 \\
2VY8 & 2GZQ & 1AMM & 1HK0 &3KZD \\
1O8X & 1ZLM & 3CO1 & 1YZM &2PPO \\
2QVK & 2X25 & 3NZL & 3HNX &1I2T \\
2LIS & 1FL0 & 3NGP & 1L3K &3ICH \\
2X5Y & 1JL1 & 1F21 & 3I35 &1BKR \\
2QVG & 1YU7 & 1HZT & 2F4K &3MSI \\
1LN4 & 2NRR & 1ZEQ & 1TZV &2YV0 \\
2RN2 & 1EW4 & 1P7S & 2ON8 &1OGW \\
2GI9 & 3EY6 & 2RB8 & 1NG6 &1X3O \\
1IGD & 2O37 & 2NR7 & 2FG1 &2FB6 \\
3EYE & 3IV4 & 2WWE & 1QAU &3DVW \\
1ULR & 1YU5 & 2V4X & 3A2Z &1NA5 \\
2JLI & 3Q6L & 3BZT & 3DFG &3KB5 \\
1G9O & 1Z21 & 3OBS & 1LMI &3BZP \\
2NT3 & 1P5F & 2FQ3 & 3BZS &3S4M \\
3GBW & 1UKF & 2VWR & 2OZF &2IWR \\
2IWN & 2GZV & 3LAX & 3A7L &2B02 \\
3I2V & 2VC8 & 1Y0M & 2PTH &2JIC \\
3I7M & 2VH7 & 2END & 1HKA &3K0N \\
3K0M & 2WLW & 2F1S & 3CTG &1XAW \\

\end{tabular}
\label{PDB_List}
\end{center}
\end{table}

\pagebreak

\begin{table}[ht]
\caption{Comparison of the average composition of the designed sequences and the natural sequences used in the parameter optimization. It is important that since we do not model Cys-Cys bond and the Proline rigid bond we have excluded them from the design alphabet. This is why the frequency associated to those amino acids is zero in the designed sequences. We are currently working on implementing such special cases in the Caterpillar model. We have highlighted in bold the amino acids types with the largest discrepancies namely: Histidine, Methionine, Tryptophan, Tyrosine. Such amino acids are know to be the one with the lowest appearance frequency in nature. Since we did not impose any restriction on the design procedure over the relative abundance of amino acids in nature it is not surprising to find the largest discrepancies in the composition for such amino acids.}
\begin{center}
\renewcommand{\arraystretch}{0.8}
\begin{tabular}{ccc}
Residue&Natural Freq& Designed Freq.\\
A & 7.23\% & 5.13\% \\ 
C & 1.08\% & 0.00\% \\ 
D & 5.92\% & 4.99\% \\ 
E & 7.00\% & 5.61\% \\ 
F & 4.12\% & 5.86\% \\ 
G & 7.55\% & 7.74\% \\ 
\textbf{H} & 2.28\% & 6.00\% \\ 
I & 5.79\% & 4.57\% \\ 
K & 6.75\% & 7.21\% \\ 
L & 8.94\% & 4.58\% \\ 
\textbf{M} & 1.72\% & 4.41\% \\ 
N & 4.27\% & 5.39\% \\ 
P & 4.28\% & 0.00\% \\ 
Q & 3.98\% & 5.62\% \\ 
R & 5.13\% & 6.73\% \\ 
S & 5.90\% & 4.85\% \\ 
T & 5.79\% & 5.25\% \\ 
V & 7.39\% & 5.14\% \\ 
\textbf{W} & 1.51\% & 4.17\% \\ 
\textbf{Y} & 3.35\% & 6.77\% \\ 

\end{tabular}
\renewcommand{\arraystretch}{1.0}
\label{Compostion_Comparison}
\end{center}
\end{table}

\pagebreak

\begin{table}[ht]
\caption{Designed sequences.}
\begin{center}
\resizebox{1.0\textwidth}{!}{
\begin{tabular}{cl}

\multirow{7}{*}{\textbf{1gab-A}} & WDDMIIRRRRFVVYYLWGSMTAEVEAEKGTNGFYYHHHDFGTKKKAQQQSNNL\tabularnewline
& WDDMIIRRRRFVVYYLWGSMTAEVEAEKGTNGFYYHHHIFGTKKKAQQQSNNL\tabularnewline
& WDDMIIRRRRIVVYYLWGSMTAEVEAEKGTNGFYYHHHFFGTKKKAQQQSNNL\tabularnewline
& WNNEVVQQQGGLLKRAARSSNDDDILHHHVFFTTKKKTEGGFYYYYRRIMMIW\tabularnewline
& WSNEVAQQQGGLLKRAARSNNDDDILHHHVFFTTKKKTEGGFYYYYRRIMMIW\tabularnewline
& WSNEVVQQQGGLLKRAAISNNDDDILHHHVFFTTKKKTEGGFYYYYRRRMMIW\tabularnewline
& WSNEVVQQQGGLLKRAARSDNDDNILHHHVFFTTKKKTEGGFYYYYRRIMMIW\tabularnewline 
\hline
\multirow{7}{*}{\textbf{1leb-A}} & AAAHHQNNNNFKKKKKKGGFQFQGGGGYYYYIEEEEIMWMFIYRHRRRRDDDDLLHVVLMWWTTTVVQSSSS\tabularnewline
& AAAVVQYGGGGGGRRVVMDWWWDIRRRYYYYNMMKEEEENNNYKAHKKKTSSQQQFFFFSDIIHHHHLLLTT\tabularnewline
& AAAVVQYGGGGVGRRVVMDWWWDIRRRYYYYNMMKEEEENNNYKAKKKKTSSQQQFFFFSDIIHHHHLLLTT\tabularnewline
& ATAVVQGGGGGGVRRRVRDWWWDDIRIYYNYNMMKAEEEYNNYYAKKKKTTSQQQFFFFSSMIHHHHLLLLT\tabularnewline
& ATAVVQYGGAGGGRRVMFFWWWIIRRRYYYYNMMKEEEENNNYKAKKKKTSSQQQVFFDSDDIHHHHLLLTT\tabularnewline
& ATAVVQYGGGGGGRRVMFFWWWIIRRRWYYYNMMKEEEENNNYKAKKKKTSSQQQVFFDSDDIHHHHLLLTT\tabularnewline
& ATAVVQYGGGGGGRRVMFFWWWIIRRRYYYTNMMKEEEENNNYKAKKKKTSSQQQVFFDSDDIHHHHLLLLT\tabularnewline 
\hline
\multirow{7}{*}{\textbf{1pou-A}} & WGWVYVVAKKKKKGNFFSSSFIIIHHHHLLLDDDDRRRRRFYYYYGGWTTTMMMEEEEEGNNNQQQQAAAT\tabularnewline
& WGWVYVVAKKKKKWNFFSSSFIIIHHHHLLLDDDDRRRRRFYYYYGGGTTTMMMEEEEEGNNNQQQQAAAT\tabularnewline
& WTWVYVVAKKKKKGNFFSSSFIIIHHHHLLLDDDDRRRRRFYYYYGGGWTTMMMEEEEEGNNNQQQQAAAT\tabularnewline
& WWSVYVVAKKKKKGNFFSSWFIIIHHHHLLLDDDDRRRRRFYYYYGGGTTTMMMEEEEEGNNNQQQQAAAT\tabularnewline
& WWSVYVVAKKKKKGNFFSWSFIIIHHHHLLLDDDDRRRRRFYYYYGGGTTTMMMEEEEEGNNNQQQQAAAT\tabularnewline
& WWWQYVVAKKKKKGNFFSSSFIIIHHHHLLLDDDDRRRRRFYYYYGGGTTTMMMEEEEEGNNNQQVQAAAT\tabularnewline
& WWWSYVVAKKKKKGNFFSSVFIIIHHHHLLLDDDDRRRRRFYYYYGGGTTTMMMEEEEEGNNNQQQQAAAT\tabularnewline 
\hline
\multirow{7}{*}{\textbf{1qyp-A}} & ADLWWWEVMMMTTTAKVGNNNQQQYYYYFFIISSSDDLLHHHFGGGEEAKKKRRRIR\tabularnewline
& AMDWWETKMMVYYVVKKKAANNTQNGGYTRILDISSSIHHHRRGGGEEQYQYFFFLL\tabularnewline
& AWWSWVVYVMMMTTTQQQQNNNLLEGGYYFIISSDDDLHHHFFGGEEAAKKKKRRRI\tabularnewline
& DAWSWVVYVMMMTTTYQQQNNNALEGGYYFIISSDDLLHHHFFGGEEAAKKKKRRRI\tabularnewline
& DDWWAVVYWMMMTTTYQQQNNNVLEGGYYFIISSSDLLHHHFFGGEEAAKKKKRRRI\tabularnewline
& DGWSMWVYVMMTTTLYQQQNNNVAEGGYYFIISDSDLLHHHFFGGEEAAKKKKRRRI\tabularnewline
& DGWWVVEGMMMTTTAKANVNNSQQQYYYYFFISSDDLLLHHHFGGGEEAKKKRRRII\tabularnewline
\hline
\multirow{7}{*}{\textbf{1sro-A}} & WEENNNQQYVGGGTRRRIIIFFGGGMMMYYYYFFHHHHHDDDISRRLLLFLQVVVTKKAAAEEQWWNTAKKKTSSS\tabularnewline
& WEENNNQQYVGGGTRRRIIIFFGGGMMMYYYYFFHHHHHDDDISRRLLLFLVVQVTKKAAAEEQWWNTAKKKTSSS\tabularnewline
& WEENNNQQYGGGTTRRRIIIFFGGGMMMYYYYFFHHHHHDDDISRRLLLFLQVVVKVKAAAEEQWWNTAKKKTSSS\tabularnewline
& WEENNNQQYGGGTTRRRIIIFFGGGMMMYYYYFFHHHHHDDDISRRLLLFLQVVVVKKAAAEEQWWNTAKKKTSSS\tabularnewline
& WEENNNQQYGGGTTRRRIIIFFGGGMMMYYYYFFHHHHHDDDISRRLLLFLVVQVVAKKAAEEQWWNTAKKKTSSS\tabularnewline
& WEENNNQQYGGGTTRRRIIIFFGGGMMMYYYYFFHHHHHDDDRSIRLLLFLQVVVVKKAAAEEQWWNTAKKKTSSS\tabularnewline
& WEENNNQQYVGGGTRRRIIIFFGGGMMMYYYYFFHHHHHDDDIRSRLLLFLQVVVTKKAAAEEQWWNTAKKKTSSS\tabularnewline 

\end{tabular}
}
\label{Seq_List_1}
\end{center}
\end{table}
\pagebreak

\begin{table}[ht]
\begin{center}
\resizebox{1.0\textwidth}{!}{
\begin{tabular}{cl}

\multirow{7}{*}{\textbf{1utg-A}} & DRIMFFFYYYYKKKKKTGFGGGGGRTTMMVWWSWVLLVVAATSSQQQQAEEEENNNNIIRRHHHHLRDDD\tabularnewline
& EITTFFYYYYYKKKKKFGFGGRRRRRHHHDDDSDLLHVVVTGSSGQAAAQQQNNNNEEIEEIMMMLWWWW\tabularnewline
& FITTIFYYYYYKKKKKFGFGGRRRRRHHHDDDSDLLHVVVTGSSGQAAAQQQNNNNEEEEEIMMMLWWWW\tabularnewline
& HRIMFFFYYYYKKKKKTTFGGGGGRTTMMVWWSWVLLVVAAGSSQQQQAEEEENNNNIIRRHRHHLDDDD\tabularnewline
& HRRMFFFYYYYKKKKKTTFGGGGGRTTMMVWWSWVLLVVAAGSSQQQQAEEEENNNNIIIRHRHHLDDDD\tabularnewline
& HRTMFFFYYYYKKKKKAGFGGGGGRRTMMVWWSWVLLVVATTSSQQQQAEEEENNNNIIIRHHHHLDDDD\tabularnewline
& HRTMFFFYYYYKKKKKAGFGGGGGRRTMMVWWSWVLLVVATTSSQQQQAEEEENNNNIIIRHRHHLDDDD\tabularnewline
\hline
\multirow{7}{*}{\textbf{1uxd-A}} & EQQQQGAEKKKKNNNNFARRARRIIDDDIHHHHYYFYYFFGGGTTTLLLSWEVVMMSWW\tabularnewline
& IAHHHYYFYYFFTRTTGRGGFQQVVVLLVQSSSGGKKKKARREEINNNLDDMAAEWDWM\tabularnewline
& IDHHHYYFYYFFTRTTGRGGFQQVVALLVQSSSGGKKKKKRREEINNNLIEWAAMMDWW\tabularnewline
& IDHHHYYFYYFFTRTTGRGGFQQVVALVLQSSSGGKKKKKRREEINNNLIEWAAMMDWW\tabularnewline
& IIAHHYYFYYFFTRTTGRGGFQQVVALVLQSSSGGKKKKKRREEINNNLDEWHAMMDWW\tabularnewline
& IIEHHYYFYYFFTRTTGRGGFQQVVALLVQSSSGGKKKKKRREHINNNLDEWAAMMDWW\tabularnewline
& IIEHHYYFYYFFTRTTGRGGFQQVVALVLQSSSGGKKKKKRREHINNNLDEWAAMMDWW\tabularnewline
\hline
\multirow{7}{*}{\textbf{1vif-A}} & WWWDMMMYYYYYQQNNAAAKKKKTTIHHHDDIISRRRRGGGGSSQVVVFFFLLLNGEEET\tabularnewline
& WWWDMMMYYYYYQQNNAAAKKKKTTIHHHIDDISRRRRGGGGSSQVVVFFFLLLNGEEET\tabularnewline
& WWWIIMMMYYYYQQNNNAAKKKKGTRRRRSDDDHHHFGGGGTSSQVVVFFLLLIAREEET\tabularnewline
& WWWIIMMMYYYYQQNNNAAKKKKTRRRRTSDDDHHHFGGGGGSSQVVVFFLLLIAREEET\tabularnewline
& WWWIIMMMYYYYQQNNNAAKKKKTRTRRRSDDDHHHFGGGGGSSQVVVFFLLLIAREEET\tabularnewline
& WWWIIMMMYYYYQQNNNAAKKKKTTRRRRDDSDHHHFGGGGGSSQVVVFFLLLIAREEET\tabularnewline
& WWWIIMMMYYYYQQNNNAAKKKKTTRRRRSDDDFHHHGGGGGSSQVVVFFLLLIAREEET\tabularnewline 
\hline
\multirow{7}{*}{\textbf{2cdx-A}} & AKKKTKSSVVVQQQNNAEEGGYFRRIIDDLLHHHFGGGEAMMMLHWWDIRRFYYYNTTKK\tabularnewline
& AKKKTSSSFVVQQQNNAEEGGYFVIIDDDLLHHHFGGGEAMMMWWWLIRRRRYYYNTKKK\tabularnewline
& AKKKTSSSVVVQQQNNAEEGGYFRRIIDDLLHHHFGGGEAMMMNHWLDIRRFYYYNTTKK\tabularnewline
& AKKKTSSSVVVQQQNNAEEGGYFRRIIDDLLHHHFGGGEAMMMWHWLDIRRFYYYNTTKK\tabularnewline
& AKKKTSSSVVVYQQNNAEEGGYFRRIIDDLLHHHFGGGEAMMMWHWLDIRRFYYYNTTKK\tabularnewline
& AKKKTSSVVVVQQQNNAEEGGYFRRIIDDLLHHHFGGGEAMMMWHWLDIRRFYYYNTTKK\tabularnewline
& ETKKFSSSNVVQQQQAEETGGYFRRIIDDLLHHHHFGGGTMMMLWWNDIRRFYYYVAKKK\tabularnewline 
\hline
\multirow{7}{*}{\textbf{2kyw-A}} & AAAEAEGYYYFHHHMIIRRRMMMTNNNNNLAKKKKTGGHHHLIRRRKKTTGGGWWWVVVVVQQQQQLLEEEGYYFFFIDDSSSSDDF\tabularnewline
& AAAEEAGYYFFHHHMIIRRRMMMTNNNNNLEKKKKTGGHHHLIRRRKKTTGGGWWWVVVVVQQQQQLLAEEGYYYFFIDDSSSSDDF\tabularnewline
& AAAEEAGYYYFHHHMIIRRRMMMTNNNNNLEKKKKTGGHHHLIRRRKKTTGGGWWWVVVVVQQQQQLLAEEGYYFFFFDDSSSSDDI\tabularnewline
& AAAEEEFYYYFHHHMIIRRRMMMTNNNNNLAKKKKTGGHHHLIRRRKKTTGGGWWWYVVVVQQQQQLLKEEGYYFGFIDDSSSSDDF\tabularnewline
& AAAEEEFYYYHHHIMIIRRRMMMTNNNNNLAKKKKTGGHHHLIRRRKKTTGGGWWWYVVVVQQQQQLLKEEGYYFFFFDDSSSSDDG\tabularnewline
& AAAEEEFYYYHHHIMIIRRRMMMTNNNNNLAKKKKTGGHHHLIRRRKKTTGGGWWWYVVVVQQQQQLLKEEGYYGFFFDDSSSSDDF\tabularnewline
& AAAEEEGFYYFHHHMIIRRRMMMTNNNNNLAKKKKTGGHHHLIRRRKKTTGGGWWWYVVVVQQQQQLLKEEGYYFFYIDDSSSSDDF\tabularnewline 

\end{tabular}
}
\label{Seq_List_2}
\end{center}
\end{table}

\pagebreak

\begin{table}[ht]
\begin{center}
\resizebox{1.0\textwidth}{!}{
\begin{tabular}{cl}

\multirow{7}{*}{\textbf{2l09-A}} & DNSSQQVLFLLTRRRRIDDNFNHHHAKKKKTKSQQVVAAGGGGGFFYYYYTIIMEEEWWWMM\tabularnewline
& DNSSQQVLFLLTRRRRIDDNHNHHHAKKKKTKSQQVVAAGGGGGFFYYYYTIIMEEEWWWMM\tabularnewline
& DNSSQQQLFLLTRRRRIDDNFNHHHAKKKKTKSQVVVAAGGGGGFFYYYYTIIEEEWWWMMM\tabularnewline
& DNSSQQVLFLLTRRRRIDDNFNHHHAKKKKTKSSQVAAVGGGGGFFYYYYTIIMEEEWWWMM\tabularnewline
& DNSSQQVLFLLTRRRRIDDNFNHHHAKKKKTKSWQVVAAGGGGGFFYYYYTIIMEEEWWQMM\tabularnewline
& DNSSQQVLFLLTRRRRIDDNHNFHHAKKKKTKSQQVVAAGGGGGFFYYYYTIIEEEWWWMMM\tabularnewline
& NDSSQQVLFLLTRRRRIDDNHNHHHAKKKKTKSQQVVAAGGGGGFFYYYYTIIMMEEWWWEM\tabularnewline 
\hline
\multirow{7}{*}{\textbf{2ptl-A}} & AMDWWSVVVDSWSSAMAAEEGGGRRRDDLLHLVYQQNNNNQQQKKKKKTTFFFFFIIHHHHYYYYMMTTEEGGGRRII\tabularnewline
& AMDWWSVVVDSWSSAMAAEEGGTRRRDDLLHLVYQQNNNNQQQKKKKKTGFFFFFIIHHHHYYYYMMTTEEGGGRRII\tabularnewline
& AMMSWWDVVASWSDDDIIRRRGGGEKAMMTYYYYYHHHHILIFFFFFGTKKKAKQNNNNQQQQVVLLSLRRTTGGEEE\tabularnewline
& AMSWWDVVASSWDLAAAEEEGGGRRRDDHLVLYYQNNNNQQQQKKKKKTTFFFFFLIHWHHYYYVMMTTEEGGGRRII\tabularnewline
& AMTSWWDVVASSSDDDIIRRRGGGLEAMMNYYYYYHHHHILFFFFFGGTKKKAKQNNNNQQQQVVLLWHRRTTGKEEE\tabularnewline
& DMSSWDVVVSSWDDAAAAEEGGGRRRIWLLMLYYQNNNNQQQQKKKKKTTFFFFGLIHHHHYYYVMMTTEEGGGRRII\tabularnewline
& DMSSWVVVASSWDDAAAEEEGGGRRRHNWLLLVLQQNNNNQQQKKKKKTTFFFHHHHIIFFYYYYMMTTEEGGGRRII\tabularnewline 
\hline
\multirow{7}{*}{\textbf{3mx7-A}} & IIRRRTTTMTLNNWWEEEYYYFHDLDLHHHHYYYMMMIIRRRIDDDSSSSFFNNNEEAAAAAKKKKKKVVWWSVVVGGGQQQQQFFGGGL\tabularnewline
& IMRRRTTTTTLLNWWEEEYYYFFDDLHHHHHYYYMMMIIRRRIDDDSSSSFFNNNEEAAAAAKKKKKKVVWWSVVVGGGQQQNQFQGGGL\tabularnewline
& IMRRRTTTTTLNLWWEEEYYYFFDDLHHHHHYYYMMMIIRRIRDDDSSSSFFNNNEEAAAAAKKKKKKVVWWSVVVGGGQQQNQFQGGGL\tabularnewline
& IMRRRTTTTTLNLWWEEEYYYFFDDLHHHHHYYYMMMIIRRRIDDDSSSSFFNNNEEAAAAAKKKKKKVVWWSVVVGGGQQQNQFQGGGL\tabularnewline
& IMRRRTTTTTLNNWWEEEYYYFHDLDLHHHHYYYMMMIIRRRIDDDSSSSFFNNNEEAAAAAKKKKKKVVWWSVVVGGGQQQQQFFGGGL\tabularnewline
& ITRRRTTTMTLNNWWEEEYYYFHDDLLHHHHYYYMMMIIRRRIDDDSSSSFFNNNEEAAAAAKKKKKKVVWWSVVVGGGQQQQQFFGGGL\tabularnewline
& ITRRRTTTMTLNNWWEEEYYYFHDLDLHHHHYYYMMMIIRRIRDDDSSSSFFNNNEEAAAAAKKKKKKVVWWSVVVGGGQQQQQFFGGGL\tabularnewline 
\hline
\multirow{7}{*}{\textbf{3nmd-E}} & AAAENNQQQEEENTTKMMYYYGGGGTKKVVVQSSSDDLLHHHRRRIIFFWW\tabularnewline
& AAAFNNQQQEEEETTTMMYYYGGGGKKKVVVNSSSDDDLHHHRRRIILFWW\tabularnewline
& AAAFNNQQQEEEETTTMMYYYGGGGKKKVVVSSSDDDLLHHHRRRIIFFWW\tabularnewline
& AAAFNNQQQEEEETTTMMYYYGGGGKKKVVVSSSDDDLLHHHRRRIIFNWW\tabularnewline
& AAAFNNQQQEEEETTTMMYYYGGGGKKKVVVSSSDDDLLHHHRRRIINFWW\tabularnewline
& AAAFNNYQQEEEETTTMMYYYWGGGKKKVVVQSSSDDLLHHHRRRIIDFNW\tabularnewline
& AAAINNQQQEEEETTKMMYYYGGGGTKKVVVQSSSDDLLHHHRRRINFFWW\tabularnewline 
\hline
\multirow{7}{*}{\textbf{3nrl-A}} & TTTTKKKKMSSEEGGGGGWWMVVVVQQQQDLLLHHHHMSIIRRRRAAAAWNNNNDDIDREEYYYYFFFF\tabularnewline
& TTTTKKKKMSSEEGGGGGWWMVVVVQQQQDLLLHHHHMSIRRRRRAAAAWNNNNDDIDIEEYYYYFFFF\tabularnewline
& TTTTKKKKMSSEEGGGGGWWWVVVVQQQQDLLLHHHHMMIIRRRRAAAASNNNNDDIDREEYYYYFFFF\tabularnewline
& TTTTKKKKMSSEEGGGGGWWWVVVVQQQQDLLLHHHHMMIRRRRRAAAASNNNNDDIDIEEYYYYFFFF\tabularnewline
& YYYYFFFFIIIHHHHGGGMMVMTRRRRRDDNNNAGGNWVVEEQQQLDDSWSLLAAAEEQWSTTTKKKKK\tabularnewline
& YYYYFFFFIIIHHHHGGGMMVMTRRRRRDDNNNEGGNWTVEEQQQLDDSWSLLAAAAEQWSTTVKKKKK\tabularnewline
& YYYYFFFFIIIHHHHGGGMMVMTRRRRRDDNNNEGGNWVEEVQQQLDDSWSLLAAAAEQWSTTTKKKKK\tabularnewline 

\end{tabular}
}
\label{Seq_List_3}
\end{center}
\end{table}
\pagebreak
\begin{table}[ht]
\begin{center}
\resizebox{1.0\textwidth}{!}{
\begin{tabular}{cl}

\multirow{7}{*}{\textbf{3nzl-A}} & AAEEQQQETTTTGGGGVLVVLSSDSDDDHHHHHISFFFFIIRRRRRVLWWWMMMYYYYYGGAKKKKKQNNNNE\tabularnewline
& AAEEQQQETTTTGGGGVVVVLSSDSDDDHHHHHIIFFFFSIRRRRRLLWWWMMMYYYYYGGAKKKKKQNNNNE\tabularnewline
& AAEEQQQETTTTGGGGVVVVLSSDSDDDHHHHHISFFFFIIRRIRRLLWWWMMMYYYYYGGAKKKKKQNNNNE\tabularnewline
& AAEEQQQETTTTGGGGVVVVLSSDSDDDHHHHHISFFFFIIRRRMRLLWWWMMRYYYYYGGAKKKKKQNNNNE\tabularnewline
& AAEEQQQETTTTGGGGVVVVLSSDSDDDHHHHLISFFFFIIRRRRRHLWWWMMMYYYYYGGAKKKKKQNNNNE\tabularnewline
& AEAEEQQYYQGGFTTGVVVHVSSDSDDDLHHHHISFFFFIIRRRRRLWWWMMMLYYTTGGAKKKKKKNNNNQE\tabularnewline
& AEAEEQQYYQGGGTGGVVVVLSSDSDDDLHHHHISFFFFIIRRRRWLWWWMMMYYYTTTGAKKKKKKNNNNQE\tabularnewline
\hline
\multirow{7}{*}{\textbf{3obh-A}} & AAQLQQVVVSSSFDDIDLHHHHGGGGGGLQATTTKKKKRRRRITNNFFNFYYYYYEEEEIMMMWWW\tabularnewline
& AAQQGQVVVSSSFDDDIHHHHKKGGGGQLLATTTTKKKRRRRIFLNFNNFYYYYYEEEEIMMMWWW\tabularnewline
& AAQQGQVVVSSSNDDDIHHHHKKGGGGQLLATTTTKKKRRRRIFLNFFNFYYYYYEEEEMMMIWWW\tabularnewline
& AAQQQAVVVSSSFDDIDHHHHKGGGGGQLLTTTTKKKKRRRRIFLNFNNFYYYYYEEEEIMMMWWW\tabularnewline
& AAQQQQFVVSSSVDDIDHHHHKGGGGGALLTTTTKKKKRRRRIFLNFNNFYYYYYEEEEIMMMWWW\tabularnewline
& AAQQQQKVVSSSFDDIDHHHHVGGGGGGLLATTTTKKKRRRRIFLNFNNFYYYYYEEEEIMMMWWW\tabularnewline
& AAQQQQVFVSSSFDDIDHHHHKGGGGGALLTTTTKKKKRRRRIVLNFNNFYYYYYEEEEIMMMWWW\tabularnewline
\hline
\multirow{7}{*}{\textbf{3obh-B}} & AAAQQQQVVVVSSSFDDDIHHHHKGFGGGGLLTTTTKKKKRRRRDFLNFNNFYYYYYEEEEIMMMWW\tabularnewline
& AAAQQQQVVVVSSSFDDIDHHHHKGRGGGGLLTTTTKKKKRRRRIFLNFNNFYYYYYEEEEMMMNWW\tabularnewline
& AAAQQQVVVFSSFFDDIDNHHHHAGGGGGGNSTTKKKKKRRRRRILLLFINFYYYYYEEETMMMWWW\tabularnewline
& AAAQQQVVVFSSFFDDIDNHHHHHGGGAGGNSTTKKKKKRRRRRILLLFINFYYYYYEEETMMMWWW\tabularnewline
& AAAQQQVVVFSSFFDDIDNHHHHHGGGEGGNSTTKKKKKRRRRRILLLFINFYYYYYWEEETMMMWW\tabularnewline
& AAAQQQVVVFSSFFDDIDNHHHHHGGGGGGNSTTKKAKKRRRRRILLLFINFYYYYYEEETMMMWWW\tabularnewline
& AAAQQQVVVFSSFFDDIDNHHHHHGGGGGGNSTTKKKKKRRRRRILLLFINFEYYYYWEEETMMMWW\tabularnewline
\hline
\multirow{7}{*}{\textbf{5icb-A}} & RWWVMMMMVGGEEGVVYQTTTGGKKKKKQQYNNNEEAAAAQSDDDNSSHHHHHIITSFFFFYYYFRRRRRIDLLL\tabularnewline
& VWWWMMMMVGGEEGVVYQTTTGGKKKKKQQYNNNEEAAAAQSDDDNSSHHHHHIITSFFFFYYYFRRRRRIDLLL\tabularnewline
& VWWWMMMMVGGEEGVYYQTTTGGKKKKKQQYNNNEEAAAAQSDDDNSSHHHHHIITSFFFFYYLFRRRRRIDLLL\tabularnewline
& WSWMMMVVEGGGGGVVYQTTTGKKKKKKQQYNNNEEEAAAAQDDDNSSHHHHHIITWFFFFYYYFRRRRIIDLLL\tabularnewline
& WVWVMMMMVGGEEGVVYQTTTGGKKKKKQQYNNNEEAAAAQSDDDNSSHHHHHIITSFFFFYYYFRRRRRIDLLL\tabularnewline
& WVWWMMMMVGGEEGVVYQTTTGGKKKKKQQYNNNEEAAAAQSDDDNSSHHHHHIITSFFFFYYYFRRRRRIDLLL\tabularnewline
& WWDVMMMMVGGEEGVVYQTTTGGKKKKKQQYNNNEEAAAAQSDDDNSSHHHHHIITSFFFFYYYFRRRRRIDLLL\tabularnewline 
\hline
\multirow{7}{*}{\textbf{5znf-A}} & AAEEGGIMLYYQQWNNTTKKRRHHVFFDDS\tabularnewline
& AAHHGGDTFFQQSSNNLKKKRREEVYYIWM\tabularnewline
& AAHHGGIFTFQQSSNNLKKKRREEVYYDWM\tabularnewline
& AAHHGGTIFFQQSSNNLKKKRREEVYYDWM\tabularnewline
& AKEEGGIMLYYQQWNNTKKARRHHVFFDDS\tabularnewline
& AMEEGGIGWLYYQQNNTRRKKKHHVVFSSD\tabularnewline
& AMKERFINNFYYSYGEGRTTKKLVHHQWSD\tabularnewline 

\end{tabular}
}
\label{Seq_List_4}
\end{center}
\end{table}

\pagebreak

\begin{table}[ht]
\caption{Sequences obtained during the last step of the matrix optimization procedure. The amino acid composition is identical for all sequences and the first is the natural sequences taken from the pdb file.}
\begin{center}
\resizebox{1.0\textwidth}{!}{
\begin{tabular}{cl}
\multirow{7}{*}{\textbf{1AMM}} & GKITFYEDRGFQGHCYECSSDCPNLQPYFSRCNSIRVDSGCWMLYERPNYQGHQYFLRRGDYPDYQQWMGFNDSIRSCRLIPQHTGTFRMRIYERDDFRGQMSEITDDCPSLQDRFHLTEVHSLNVLEGSWVLYEMPSYRGRQYLLRPGEYRRYLDWGAMNAKVGSLRRVMDFY \tabularnewline
& GRIQCGEDRYGQGMCNIYSMNYDPRMPERFRCRSCDDDIGLWEYYEQRNYRFHQLNLIRGRYPDQYGWCYSLYELVCRFLQPLNPSTSMMRRQYFDFRRGTVSEMTLLRPVIYDRFHPTGVHPQQKDWGDLVCDELVSYRGNKYLSTSQESSRFYSWGAGHAIHGFDRLDMFSE \tabularnewline
& GLITMYYDRVFYSRCMLEDGQQELLIDHFSKCDSYYVPSHTWMMYEFLNYQGFQHFPTPGNEPNLQDYDSDNEQVLYCFYSCLHTGTARSRIFERDDFRGYQREIRSDCLRWSGRRRIKERASLGELCGSWVNQSMPYQGCRLYLPRPGDYIRRPDWGGMNHGVYSRQRVMDFR \tabularnewline
& MYITFYEDRGYQGVCYECSSDKYGILVIGSRCYSYRVDSGTWMLSCVLQHQFHQYFFRRYSYPSYRQWMVMHCNDDPDQLPPGNTGRFRMRIEDRDDFRGLNSEIEDDRPSRQLSRHLTECHSQKGLEGFWNQYEMPLYRGCQYELTPGFDRLGLIWNARVARNGSLRRPMDFY \tabularnewline
& GFYTFEEKRGFQPHCPPCSSDGVLLQPLFSRCNSIRVQSGRGVDYEEELYWYRQYFLRCMLYQDYGDKMLFYDSQTNCRNMPQHTYTFRMRIWEGDDWGRQMSGISDDCPSLQDRPHLQEPHSLNVCRTSGVLYEMFSYRGNIGLLRYGRYSRYIRIEAMNADRRDHRWVGDFY \tabularnewline
& YPYIDHSWRGEQGFCCYGLSWMYLLQGGSQRCKFIRERLLCQMADDRPRDDQYWTMCSRSDLFGYQSWMGTDDRTRSFRRVPVHMSTFRSLIYNVRDSELIYQEIYFDVPQYQNYHDGTMVSFLNHHPDLGLGSEEYGCRGYQYGNRPSEYRFPLFNPARNIKRGERERVMDLC \tabularnewline
& QRIDFGEDERRQSYNRFSSSRYPLRLFYMGRCNQIYWDRSCQMLIEFWNNIGHQYVLHRGDRNDYQMWEGIRDSKLSCGLIPNYMRTQRFRVRRRHCYDPQTSCLTYDCPSLSDLFYGWKVHSDRVLEGYTERAEQPSGPLTEPVLYPGRHMFFLDYDAMQYEGGSFVGGMDCY \tabularnewline
\hline
\multirow{7}{*}{\textbf{1BKR}} & KSAKDALLLWCQMKTAGYPNVNIHNFTTSWRDGMAFNALIHKHRPDLIDFDKLKKSNAHYNLQNAFNLAEQHLGLTKLLDPEDISVDHPDEKSIITYVVTYYHYFSKM \tabularnewline
& TSGADLIKLTINFKTDGYFKDDIKNKTMPARDNQVMKPAMALHRASLLSFQNLLLVNPEYNHCLWWFDADHHNLNFHDLHAYYKSVIKYEELSAPQDKHTKYGVTSII \tabularnewline
& IKGKDALLHFCKNQTKIYPPHDGAKNIPSKRDAMKFMLLYHQWDETEKDADPGKKSNAFYTLQLDASLFLHHIENSNLDTIVRANHWDVIALLTLTYVVNYNHYFSSM \tabularnewline
& TSGADLIKLTINFKTDGYFKHDIKMKAMPARDNQVKKPANALHKTSLISFQNLLLVNPEYDDCLWWFNHNHHYLDFHDLAAYNMSVIRYEELSAPQDKHTKYGVTSLI \tabularnewline
& DSWKKALLYWHKNKTAGYPYKNHNNFTTSARLGDSLAAVIHFHNPMDFSLDKKNLSAAHLNLHLAFNLIEQQQGVTKLDDPEDIIVLHPMEKSIITYDVTYYMDFRKC \tabularnewline
& ASASWKLLAWKQMASAGYKEVNHLLITTSIHDRMAFNNLIDKHERDLAYYDDKLKFNFPYTIQNEFIDDHQHYGKTCLLPPNLFSVDLPHAKSLITGVVTDNHYNKKM \tabularnewline
& KSALDAILWWDDVKTMLYPNSNLHQFLVDLRPAAAFKALKYTHCPDKILFSKKKLSDKHDNLNNGHNRFEQHLGGYKYVNIEYISVDNPDETIQITTTLAMAHYFSHM \tabularnewline
\hline
\multirow{7}{*}{\textbf{1EW4}} & MNDSEFHRLADQLWLTIEERLDDWDGDSDIDCEINGGVLTITFENGSKIIINRQEPLHQVWLATKQGGYHFDLKGDEWICDRSGETFWDLLEQAATQQAGETVSFR \tabularnewline
& GTATERFHSSAVDGQFIQIRNLRWNLGDLDKCLIQETLWLLEKDFGLVIIIATWDHEDQDGTLDEATLIHNYGFEEEWMDQDSGEQSWCQGNPARKIDFEGTDSRV \tabularnewline
& MTATEVFHSSAVDGGFIQGRNLRWQLHDLLKCEIQETDRLLLKDFGLVIIIATWDIEDQDGTLDEATLIHNYGFEEEWGDNDSGEQWWCQQNPARKIDFEGTDSSR \tabularnewline
& MNQSWFHDDADGVWIFIEERLGDWDGDSDIDQEGNDGRLTITWEGSSTIIVNGQEQLWQVELFTKCDGYHFDLHGDEELCLLSANTFKRLLEQIATPQTIEKARAR \tabularnewline
& SWCRQDMHLEDWWWLCLEALQLRNDILGDIFDALGDDSLGRHQKIETNIDSDVETPITAVLVQGDEINDNKTAQGGTEQEGHYQWTGIRFEAKLSTIFFDEERGFS \tabularnewline
& ANDSDHWRLPSELLQTIEQFHDDLDGTSAYDQMINWTCFRETQEKGGIEIIERRDNIHLVWLFTKEWDGQLDIGGEEIVSDDCKQRLTVLLNQGADEAAGFTWSFG \tabularnewline
& ANDSDHWRLPCELLQTIEQFLDDKDGDSAYDGQINWTVFRETQEKGGIEIINRWDMDHLVWLFTKFSVGCLDINGEEIIRDGSHERLGDLLEQAATQQAGETWSFT \tabularnewline
\hline
\multirow{7}{*}{\textbf{1F21}} & KQVEIFTDGSALGNPGPGGYGAILRYRGREKTFSAGYTRTTNNRMELMAAIVALEALKEHAEVILSTDSQYVRQGITQWIHNWKKRGWKTADKKPVKNVDLWQRLDAALGQHQIKWEWVKGHAGHPENERADELARAAAMNPTLEDTGYQVE \tabularnewline
& WQYEHFTGHSGLGNPGPGGWLAIARYSKAEKDVRDTYTRTTDNNMPLMAANGAQEHLLHILGVLAETASAYVRQEPTKYIEIWQVVEGAGAMAKTVRKVDNWWRKDGLLTQKQIKNEWFQGHGDLAEIARKPERKRAAGKLNISEDTARQVE \tabularnewline
& GQAETHTRDSGLHNRGWGVAISIPLYTHREKNATKDLKRTSENAVELMKGNVAYVAQKEDAEVLARTDMEYWRQGIQQLIYWWGPWGWGILDPSLMKADPALLRVFTALHAHVAKAEQGGGNAGGQENERIYEKRRPATTNKATEDKKFQVI \tabularnewline
& AGETEFVQTVAATNAGPYADGAIRALRGVIKTMKRGYLRIDKINIISMAKTHADEARDGVVKKKRHTEFNYLGQQILEGQQNWQSVYWWLKPDHYAANKSLWRDLLRVTGQLQGEDEWRHGTPLKPHEESAGENATAAALNPMWETRGGKEV \tabularnewline
& KWYKMATLLLAKGTLGPGQYQARVQTNGTEKTFAVYIMVGTPHKRNLQAAIWAGEVETKHIQVPLDTDMIQEEEGRFRGIQYWVARKWKTGTGKVEKNWAAAGRLEWKLSANIREALHVSGHSDLDAHIDDDRYNRAAARNPNGESEGQLPE \tabularnewline
& IQVDLFTDGSAIGKPNVGKLAAIDVGRWRRARFGAQWDARATNTMHLMAPRTAEGELNEKNTPIESGTSQEVAARIAEWIGYWKKYSWVGGGKGEVDVALTEGYLPRATGQHEIKWHLLKQHAREDHNQYQDKKKRAALMNPNLTETLYQVE \tabularnewline
& KATGGFVEGLAASDDGMKLYIWNTIIRHDTWKFKAAYYTKTNNGEKQMDAGVAGTEGTEHLKVWESRLSQRVRQGIKRAHHNRIKEEVNLSDDQPWDNVILKPRQGAALMEQLITWKWEHPAPEAPRNTYVAERGRGGALLQLGAATEYQVA \tabularnewline
\hline
\multirow{7}{*}{\textbf{1FL0}} & IDVSRLDLRIGCIITARKHPDADSLYVEEVDVGEIAPRTVVSGLVNHVPLEQMQNRMVILLCNLKPAKMRGVLSQAMVMCASSPEKIEILAPPNGSVPGDRITFDAFPGEPDKELNPKKKIWEQIQPDLHTNDECVATYKGVPFEVKGKGVCRAQTMSNSGIKL \tabularnewline
& NDDSRPDYRIGTMSIAVVNEDADSLYVEEPKVIEVAPRLRVSLLVHQVPVEQMQNRMKCLPCLKKMAKDRHMLIDKMADCVSSTEKRTILAPLIGSVNGAGITFIVFPGECPKELVKKKLGWEQTQPPIHPNQEDIATLKGVPFENPGAGNCRAGVGSVSIIDL \tabularnewline
& NDDSWPDYRIGDASIAVGNEDADSLYVEEPKVIEVAPRLRVSKLVHQVPVEQMQNRMKCLPCLKKMAKTRHMMIDKMADCVSSTEKRTDLAPNIGSVLGAGITFIVFPGECPKELVKKLLGREQTQPPIHKNQEIIATLPVVPFENPGLGNCRAGVGSVSIIDL \tabularnewline
& NDDSWPDYRIGDASIAVGNEDAPSLHVEEPKIIEVAPRLRVSKLVHLVPVEQMQNRMKCGPCLKKMKATRYMMSDKMSDCVSSTEKRTDLADNIGTVLGAGITFIVFPGECPKELVKKLVGREQTQPPVHKNQEIIAILPVLPFENPGQGNCRAGVLAVSIIDL \tabularnewline
& IDVSRAALETGKPGMNEGHLDFTSCYVTEVDVVVIPIRKRVSGLENTGQVEQGQEMMIILLCNLLVAKMRMVWSQGKGRCASSPSKAEVLVPPIASVLGDPSTGDAHVNEPDKEINDIKKALIQLQPDPATKDICVAIIRGVPFEPKDRFLCRNPHMYNPKEKL \tabularnewline
& IDVIRGALETGKPGMNEGHLDFTSCYVTEVDVVVIPICKRVFGLEITGQVEQGQEMMIILLCNLLVAKMRAVWSQGKARCGSSPSKAEVLVPPIASVLGDPSTADAHVNEPDKEINDSKKALNQLQPRVATKDIRVMIIRGPPFEPKDRSLCDNPHMYNPKEKL \tabularnewline
& AEVPRVTLDIGEIITNVPMPDKQSLSVEELDVFEIARRSVVSRLDNHLLLEMMQNEHRILRSNRGAAPMGETPVQPMVAKARSKEYIPGLAPKNLSLPCDSPTFDMCPGAIDKDGAIKKKIWEQIQGGVHYNVICPVKVKGVPGCIKDCDVGLAQTFSNTEVKL \tabularnewline
\hline
\multirow{7}{*}{\textbf{1G9O}} & RMLPRLCCLEKGPNGYGFHLHGEKGKLGQYIRLVEPGSPAEKAGLLAGDRLVEVNGENVEKETHQQVVSRIRAALNAVRLLVVDPETDEQL \tabularnewline
& GLLQKEKSERNEMNERGVHLVGVGSKLALAIVLATVPGRERPEFACYLKCLPVLTGGPNLAEVHGKEYGALRLRQDIDGEDLVPHEQNVQR \tabularnewline
& FVEDVHGVLGLQGGVNLGEPHPARGVLGKTICLANNPELYGPEGLYCRSLKGLKRNEDKEDEEREQVSRAAHILLVVMLLVKARPRTEQQA \tabularnewline
& RMLTVLCCKELPFEGYDPHLHGETQVLGQLNRLYKPGLPAGKAGKKAGDREVEVNNENVSLEEHGQVSVRIRAALIAVRLLVRDPEGGEQL \tabularnewline
& ILLPRLNPQEAGESPYVFGLVGEKGGLRRYLLAHEEHNNAEKKGVQKGDGLLLIPGECVDKMTPNQVVEGLQAVLRAARRLVHECETSDVR \tabularnewline
& LMVHQLCNLLKNPGGQEVLALLCAPEYIQVPDNVGGASIAEKGDRGVRGLGVHKARLNLEEGTEHYAEKERRKQRVFVLLSVREPPTGEDL \tabularnewline
& RNSPRVCEDEPGYKALLFHGHGVKKYVGNPILDEEVRTNAEGAGLVLGEHAVELPGENVRKKGRQQVLLRSGCELQAQPLLEVLITADRML \tabularnewline
\hline
\multirow{7}{*}{\textbf{1HK0}} & GKITLYEDRGFQGRHYECSSDHPNLQPYLSRCNSARVDSGCWMLYEQPNYSGLQYFLRRGDYADHQQWMGLSDSVRSCRLIPHSGSHRIRLYEREDYRGQMIEFTEDCSCLQDRFRFNEIHSLNVLEGSWVLYELSNYRGRQYLLMPGDYRRYQDWGATNARVGSLRRVIDFS \tabularnewline
& GGITNYMDRLFSLRGWEELRDHPRGQWMGRRCMRFRHDSGCWSCLEQINYSRLQYFLEPGDSSDVQEWYGASDQQRFCNLIPHYVSYRIGLYERSNIRSLRRQGTEDLSCLQHEFRSLELHSLNVRRCGPYLVVAMNYSYSFYYLSPKDGRHYQDSGATRQNVDLGIEYDDAR \tabularnewline
& GHITRLERDGFRGRRIECYSQHSSQRYYRGRNHSCRADSYCRMNYMQPNRFGSQRFDGRGDSTDVWMLMSSNPSWWECQYYPHAQSHRINLYERWDVQGQAIESCPALFHLRYEEKFDFGRSDQILLVGVVDYGESNYELRTYLNSGGDLLRSRDYCDVLLEYLSLRLQGILP \tabularnewline
& GRITRLERQGFRGRRIECYSQHENQRYYRGRSHSCRADSYCRMLYMQPNVFGSQRFDGRGDSTDVWMLMSSNPSWWSCQNYPHADSHRINLYERWDRQGQAIESCPALFHLRYEEKFDFIHSDQGLLVGVVDYGESNYELRTYLYSGGDLLRSRDYCGVNLEYLSLRLQDILP \tabularnewline
& EPIWSYLSRGYIGLLEGCDFSHSDEQPYRNRCNSQEMNSGYWVHYECLLSYWCAPRRRHYRYTLSQTGDSLSDSVRLRRLPAHCDSHRIRRDEFQGGLDYHIEGVYFQSTLQCLFRFNGIMYRFVELEDRNLYEDQNQSRRQGLKMPGGYGRYQSWGALMDRVDALSRVIDNS \tabularnewline
& GHLDILVYRQFPSPHCEHSSGLYDLHQPLSSCNRARVDSRSHSIYEQGNLEGLQRGIRTVDYERWQQWMGGSDLMRDCRLIRHPFDGRIKLYEDERDRQQMIECLCYCSSLGRSFWFNEYYRYNGLEFSRSLYALQNYRDGQYVLMPSNYDFYSDRWVTNARAGGGRRVLTES \tabularnewline
& MEITLLECRIFQGESQEVSSDLSSFLPYLSLHNSARIDDGCWSKYLHPNLYGDQNFYYRWSYVRHQQGRRLSDSCRSCAHEPGSGEHRIRLYISERYRGQMLNFYYDCGDEQSDFDNLERRSMFGVCWPWVYYHRGNYRGMRYELQLGDTGLRQVRQATNDGLRPLRRVADIS \tabularnewline
\hline
\multirow{7}{*}{\textbf{1HKA}} & TVAYIAIGSNLASPLEQVNAALKALGDIPESHILTVSSFYRTPPLGPQDQPDYLNAAVALETSLAPEELLNHTQRIELQQGRVRKAERWGPRTLDLDIMLFGNEVINTERLTVPHYDMKNRGFMLWPLFEIAPELVFPDGEMLRQILHTRAFDKLNKW \tabularnewline
& TVAAMEGGGILLIPLELSNAALKRLRWKVDSFIILVPSFYPDPTNRRTHQPDYLNQHVTLTLILAPVELYAHNQRWEEQVPILAIAERTGPRSQSDDIMPFGNEEDNPARKRKLHWQLLLRVDMILPNFSEAGENVLPTGEMLDQLFATAAFEKLTGY \tabularnewline
& TVAFIAIGSNQASPLEQLIAAKRAFGDIPENYNLAVSSFYRTPPLGPPDQQDYLNTAVALETSLAEEELFNHRARIELQQGRVRKLREWGPRTRDLTIMLVGNEFIPLSRQTVPHYAMKPKGFMLWPLHEIANELKLPDGEMDTLDLHVLDTIVLNLW \tabularnewline
& TPEKFANLMDETAPLRLVNAALGALGRPPQSHINTVSIFYLKSELGRTFQPDYQWAAVTDERRKAPEEPNNHTHEIVLQQSGNRIAELLGNNTRDGDYMLITLPIGVPEQLSVPHYIMKRRDDMLWLLIEIAPALVFVDGEQLARSLETLAFFKPLLW \tabularnewline
& APAYAIILSREFKLQLGANNAEKGPEHPQPSLENLDQSFYGTRLGHSQDLMLLPAAIEPMETIAAPELLLLDDVRILLQQSRVTSELTWGRVTGLLDIMLEVNEEINTTRVKVFHYDAKNWGPYLRFDAVIAPERPFPWNKMARQTLGIRLFDTPVNH \tabularnewline
& EVAIIVIPSLLASPDDQTRATAVALLDPPEQHSLTVSSMLRTYALRSEDQPDDPRYALFLNFVLDPLTLLKEKQNYEHAQDTANGGEMWGPEVLNLRIMPWKNERIITARLKRPHYIGNNRGFFIVPLLHIAAELWFEPGGMLTQFLTERVQENLAKG \tabularnewline
& RVQYVEASPPVASPLQRLNAPTAPLEGIPYFHITTLSSQVELPATLNQAQWDSLVLAGELRYRGDSEEGLAKLLTTTRFALRLRKAERMHPPILDQDIMANGIAKLTLFGVGLPQYHNNNNHFWLPPIKEVAVELIDEDDEMLNRILMIRFTDKGWFE \tabularnewline
\end{tabular}
}
\label{Seq_List_5}
\end{center}
\end{table}

\begin{table}[ht]
\begin{center}
\resizebox{1.0\textwidth}{!}{
\begin{tabular}{cl}
\multirow{7}{*}{\textbf{1HZT}} & LHLAFSSWLFNAKGQLLVTRRALSKKAWPGVWTNSVCGHPQLGESNEDAVIRRCRYELGVEITPPESIYPDFRYRATDPSGIVENEVCPVFAARTTSALQINDDEVMDYQWCDLADVLHGIDATPWAFSPWMVMQATNREARKRLSAFTQLKL \tabularnewline
& LHLPFMKWDHNAPRTALSAVQALSWPPWFVNACNKICAEACLGEKDERWVDDTRTYEQGLEVDGADTRLPRTTYAGRGSSGHVECSAVTKFAASPFSLDQVPMIGVFVIQWQRFAYVLESYPVLPWNDSANMLLTKLNDEIRERLIIATQRSR \tabularnewline
& AHVSFSSCIFNAKAQLDVEPRALSKGLSPGDWTNAAWGHPAAGESNEFARCRKCERELGQEITPREGIYPDWRVDATMPLKQLLNEFCQVFAARTTSLLVIKDDEVMDYYIWLLDLVLHAIPVTAFAVSSWMVLQQTNRTDYNRRPWGPVSRT \tabularnewline
& LPAGFSLLTHLAKYCWMVKYGALNKEAWPDWFTNPHFGFVELVLATEDAVITRNREEQGSEIRPSSCTWPIFYMRAIPQFDEVSAVSCPVWAHRQTLAWSLVITENLDVPGCDLASYNAAIDQVPDNTKSDGSMLRRVRQARQREDRATGLKL \tabularnewline
& PFLAFPFRLDSADAQALVTVLRESGVSWIGYWTNGSTISASEKLENELAVAREERYPIDAKLNPQLSDGCERYPVLTMFSGIKQMVVVPLCAARLTANLYQNLRSHMARQWFDNVWIPHADDVTPWTFEVRRAIPGLCRWGDKEHATQTSCKD \tabularnewline
& TLLAAASSGTVIYGKLALTRKASVVESCPGVATYSFALHNQTRPSEWDFSIRRARCEDVVPCTLSVRNINEFRKRLVWRGKEPMNNDIPLQDSLTVDAPTGNDDQFMDYQWCDFHGVLLGILARPWAWLPFMHAIETPREAAYVLSAWQEQKE \tabularnewline
& AASAFLSEEELDKGTFLVGNRLRSFDCDILSWRQSRPDTNQCGCLIGEAQSFELRYWVGVALTPPEFQYSALWERAMDVTGIVPNNVVPYIAANTVSTHQIPDIRKAMRPRDKCADVRVLWAMTPWAETPWKLLLENVDHARHTLSAFSQYKG \tabularnewline
\hline
\multirow{7}{*}{\textbf{1I2T}} & HRQALGERLYPRVQAMQPAFASKITGMLLELSPAQLLLLLASEDSLRARVDEAMELIIAHG \tabularnewline
& ARLLGAMLQTPMVLDRQLVAQQLHIDMFKEALARAPLSRSASYEEEAGRLHSLIELPILAG \tabularnewline
& RPQALIERMSRFVQAMQPARAIALTGAKLELSVLDLLYGALSEDSLHARPGEAMELILLHQ \tabularnewline
& RLAAVLERKYPTLQMMQLAESMVIIGHLEALFPILDLRGLASQDLLSGHQAEASEAARLRP \tabularnewline
& HMPGLRARAAIVELFRLQMSLHLPMQSQAEEIKAGLLLGLLRRDSEPILVAAEAQTSLADY \tabularnewline
& HDVILGLFLLEHPRAMQSPAAVLTRSMARELERLQIIALALRESGLLMLPYDKEQAGAQSA \tabularnewline
& MPGSDASRELLAVLALAQMFDLQSLMEAAAESPHLLLRVLEIIRAQPQGRAHLGRKYLTEI \tabularnewline
\hline
\multirow{7}{*}{\textbf{1IFC}} & AFDGTWKVDRNENYEKFMEKMGINVVKRKLGAHDNLKLTITQEGNKFTVKESSNFRNIDVVFELGVDFAYSLADGTELTGTWTMEGNKLVGKFKRVDNGKELIAVREISGNELIQTYTYEGVEAKRIFKKE \tabularnewline
& NFDEFWGVDRGENTESVSMTTKGALVVRLFGIHSNIEYKEYGEENKKLVLFDVNSRTLFLMRWKKYKEKVNNAGDLKGEKIDTMVKGKNLEDIFGGDNQKNVIAEVTIAGVELKQTATYGFRTAKRIFETE \tabularnewline
& LENGGAGEDLNSEYLRFMEEMFKNTVNLKLYVYLNRIFEIHQEMVNFADKATLRFDLKVAVIERETGKVTSTKNGTTGEITFSRSGKDVWKKVLKNDFKKGAGGNTEIVKREDGIQNAVKGVEWYDTEIKF \tabularnewline
& NFDEFWDVGRGENTESVSMTTKGAAVVRLFGGGSNIEYKEYHEENKKLVLFDEDSRTLFLMRWKKYKEKVNNAGDTKGEKINLMDKIKNLEDIFGGVNQKNVIAAVTIEGVELKQTATYGFRTLKRIFETV \tabularnewline
& VGDGTDKVKRNELDEKFRKTTFLNEVKMDLRAHDNLKLKITQTGNKFVVNESSNNNDIWVLTVFGVAFRFSLADKTALTGGWFMETNKLVGKIKRVENGKEIIAVRMISGKEEYQTYEYEGEEAGGIFKYE \tabularnewline
& ALDGMDKVDRNENGITVEEKTNNNVEKRILEAWDNEELTNTQKGFKFTVKKWNIFRINDVVFGLGHDFAYSLAKKFTLTGTSTKEGNALVGMSKRVDSGYGLKEKAVIFGEVLIRTYEYMGVEQKRIFKEE \tabularnewline
& AFIGIMVVNTTENVSLFIEKMGVDLKVTDLDKDVVKKWGFNQEGNENTLKGISTGRGVFVRFELGKYFAGSELTRTTLEGNKFMENVKGKYKITRKDADEEQEAIRVISNWKLKNNTEYEYAEALGRFHKD \tabularnewline
\hline
\multirow{7}{*}{\textbf{1IGD}} & MTPAVTTYKLVINGKTLKGETTTKAVDAETAEKAFKQYANDNGVDGVWTYDDATKTFTVTE \tabularnewline
& GADVTGTEDLVTADYELAKTEFKGGTETNTDNWTDPQKTNKVVKTAAIKVYAYTFTTAMKV \tabularnewline
& YTTETTTPELETKFYKAQTKYTTDKGKVANNAFTGKINVAVDAATVAVMDETGGTLWDVDK \tabularnewline
& NGGKAELVTTKTYKATAGDTKTVTVVEELFTVDNTKAYIMNTGDWKQTETAAYDFVTPDKA \tabularnewline
& DTPTTTAYAKVKKDTNKTTKEYADKIDAATQWEEDVNMKFAAETNVVVTLFTVTYLGGGGT \tabularnewline
& IEWTAGGYTLAAVDDTGEKNKTTTAKDAKTGYKEETNYLVKMTTTTVPQDDTFNAKFAVVV \tabularnewline
& ANTAKAKYGFDLAVTMLTATKAFKYEEDETQTKAVVGTTINGGTPNVVDVKETTTTKWDDY \tabularnewline
\hline
\multirow{7}{*}{\textbf{1JL1}} & KQVEIFTAGSALGNPGPGGYGAILRYRGREKTFSAGYTRTTNNRMELMAAIVALEALKEHAEVILSTDSQYVRQGITQWIHNWKKRGWKTADKKPVKNVDLWQRLDAALGQHQIKWEWVKGHAGHPENERADELARAAAMNPTLEDTGYQVE \tabularnewline
& KQVEMQGAANRLSNKYKRPDGAIVHARGLGETFGAGKIATTNTPMTGMAQIVALLPTAEDAALIYEADDGKRHQSATQWDNKWGVRLQQTYINEPHNNVDAWQRGYAAFAEPVVEWSWLGLKKHGLEEGRVEWLAREIRLKGTIRHTEYKKS \tabularnewline
& KQVALQGAANRLGNKYKRPDGAIVHARGLGETFSAGKIATTNTPMTGMAQIVALLETAEDAELILEKDDGKNHQSATQWDRKWGVRYQQTYINEPHNNVDPWQRGYAAFAEPVVLWSWEGLKAHGLEEGRVEWLARAIRMKGTIRHTEYKKS \tabularnewline
& KQVRLQGAANRLGNKYKRPDGAIVHARGGGETFSAGEVATTNTPMALMEQIVALLATAEHAELILEKDDGKNHQSATQWDRNWGVRYQQTYINEPHKNVDPWQRGYAAKAEPVILWSWEGLKAGHLEEGRVEWLARAITMKGTIRDTKYFKS \tabularnewline
& WLLLIATAYSALKMGKKGGNGQHVQTRGKMGEFRAIWTRQTKWSLALMAGGTTLEAVEVPIEEDQRTDWQRYHATIAYWKNSVEKKGENAANAIPVRGRILYLFAQDETNDGPIPKEAKSGHALHPDEERAGKLTRQAYNQARWNDVGVVHE \tabularnewline
& KSKEADPHTGKLWMWRPLGYGYIGRYKGRTHPATSGALRTEYTGSQLAAVFAAKVRGEGKEVNILLGDSQYAAENEANTQAIIPKRHLEANNGRAVKNTDKWEREDLAQGQEATKNKWQMIHTDHMERLVTDALARIAVIWPVLEGQGWQVF \tabularnewline
& KDEYLVTAGENLAIVKPDIGAQGEGAHNAWGGYSVKYTFKVYWQFELTAAIGLAERAKESAAEQENTATAYREKAIQQGIHGWWARTGTTDRVKISKSVALLQRTRDGLNQHAIKWEWPKTDMKVPRNHRMLNLARPQGLVPLNEDMGRGHE \tabularnewline
\hline
\multirow{7}{*}{\textbf{1L3K}} & KEPEQLRKLFIGGLSFETTDESLRSHFEQWGTLTDCVVMRDPNTKRSRGFGFVTYATVEEVDAAMNARPHKVDGRVVEPKRAVSTVKKIFVGGIKEDTEEHHLRDYFEQYGKIEVIEIMTDRGSGKKRGFAFVTFDDHDSVDKIVIQKYHTVNGHNCEVRKAL \tabularnewline
& ETPFTLDILDEKGLSFQTTETMLKSHDKQWGGLFPCNVMRDPCISKSRSFGFVRYATEDVYTAHSKMTKHGVTERVAGPIVNVHDVKKIVVEGLGTRNIELEHADYFGVKFRIEDIKDATGEFRGKKRGFAFVEKDDRVSRDEIEVQVYHHVTGENEQVRARK \tabularnewline
& GAKKLQRKHMIQKMSPETPRESQWISHNDGGRLTDCLGRKVSVDTGSLGEGFRTDALKRVVFEAMNVNPFEPDIRVDVVKTRIRGIHKIFEEEKKHTDEAVYERHYFGDYVEIEEVGIFTSTVSGKKAGVAFDFFDDKENVDTTLGKTVFTVRYHECHVLQAR \tabularnewline
& GKPESHRKGNNMSLGVPTDDRKDRGLVGVWRKLVGEIEAEVPVKIGSTRFAFHTFGTGEAIDNNMIDAPHKFDSRGGKKCVACQFIQKKFYHEVTTEEHVAELRDDIVQKGMREFKEIHTDRESVTKRDFAVVLRYDTTSETSYVIQLYGTVFVHEVEDRKFL \tabularnewline
& KEDEQLAKLFIGILSKPTTDEQGFSHDGQWGTLTPCVVGVDENTKRFRRFLFVREATEEETDAANMARPHKTDERVVHPGRAVSTVDHIFVGEIKYDTEEKHLRDYFEQYSKIKVINGMTKRGSGGKRMSGRVVFDKHDSEDKIVIFYFHTVEGVNCVVRKAL \tabularnewline
& VYVEQLGVLFRGRDHNAQKGIFIREHSTNWGTLMQFTGGHDPNHKRSFGFKTTTVMTAERVVKAVRRTVDKAYDNSRFPVAARKVVISIEVGGIDEDLFEHPKRTELKAQGKKEKICIMFGEDTYKERGPEFVTFDDHDEVELVVESKGHDVDSSRCEYTKIL \tabularnewline
& KEPGHARKTEIKDLSETTTEVRLHVDVEQIGELTCCVDGRDMNDARSSGVGQWTAPTFELVRNTYTKKPDVVDGMVVEAKRVFSLVKKGGVGEIDEDYEFQKLRFQFFEYKKIAVHEAMTDRHHKGRRIFIFTNFVDGSSEDFGVIYKPHTVNGHLKRSRIAE \tabularnewline
\hline
\multirow{7}{*}{\textbf{1LMI}} & SAYPITGKLGSELTMTDTVGQVVLGWKVSDLKSSTAVIPGYPVAGQVWEATATVNAIRGSVTPAVSQFNARTADGINYRVLWQAAGPDTISGATIPQGEQSTGKIYFDVTGPSPTIVAMNNGMEDLLIWEP \tabularnewline
& SAYMITGKLGPETKAADLVGQVVLGWKVSDFKSSTAVEVGEMVASQVWENPATTNTNRGSVTPASSQTNIRTMGGITPRVDWQAAGVDPISTAGILQGEQSTGYIYFDVTGPDPTIVPAIIGAYLLLPWAN \tabularnewline
& SAYEMRGTYGSDGLELDGVGIVPWDWYTISPPAVTAITVGKGAAEVVLEAIATPVGSVASGTGSNSQFTALTAFTNNLRVLWQAAGPPVDSGKTVNQQWQSTPKIKIDRVGPSTQYTTMNPIMEDALIVIG \tabularnewline
& AIYSQTGLLDSTGYVEIVVAILVPDWPFKDELFKTAAIPIRPMQAMISTITQSVPQPGTSVTPGGAAVNAYEVEGANVMGQNVATSGTIVWQATKPTKESSIGTTNRDGSGARPWAVSDTWGLVDLNYGGL \tabularnewline
& SPQPSALKLENRMGLTLTAGTTVNVIKWSGGYWGQAVFDDYPSDGEVQGATDAVVDGFGSYTPPGSVVLVESARIWNVTRSTIAWGQYTLSKANPPQGIQITAGISLEVTTPVGAITAMKNTMEADIPAVI \tabularnewline
& TWYFIPPDVGVSWTGTNTGGLFVQDRKVSPLYGSVGDASTIIQTASVWESAAAGQLIAISGTLTYKIVNVRTAVPVNGISNWAADVDNTTDGQGGPKTELSYQEQIGPATGMRAASVAMIKPMEVPLTLEP \tabularnewline
& NIYIDTWSLDTPLVMPSYRFDGNTVWEVSGIPSSTAVTPGTKVQGSGQKDGAIVNGARGKEILAVSPVTNTGAMVTASPESWAQVQAATTSTDYIRQGYAITGKVALAVVGPFPTIEADNGGMEQLLIWLP \tabularnewline
\hline
\multirow{7}{*}{\textbf{1LN4}} & MDLSTKQKQHLKGLAHPLKPVVLLGSNGLTEGVLAEIEQALEHHELIKVKIATEDRETKTLIVEAIVRETGACNVQVIGKTLVLYRPTKERKISLPLE \tabularnewline
& IELSGDMPQVLLLTTVTLKKLLLHHLTEHEVGLKPVVKIHIELEAKKLTDALAECRRVGEKQAEAIRVETSPQGRKNTGIVLPAYKIQTEIKNSVGLE \tabularnewline
& HQLSTKVKTLLEGHAEELVLVTETLLIGAKVGKEALMTLILTVIEVKHNKRIPPDQGKSQIEVDKGEPEREACARPAIEKTTVHNRLVLLQKYSIGLL \tabularnewline
& MRLLTLRLQHLKGKLHPLQKVVELGSNGLTIGVSNHIEQELEHEYLIKEKIATEDREVKTEETVAIVDETSACAVKVTAPTAVLKQPIKGRKIGLPLL \tabularnewline
& ENLSTAKKLYLTVPRKIQLQERLAGETRPITIVKLETEQHLLLKAGIVVAVGSLQGHLKTKIEVELMIDVHHCDGVVIKKLLGPTRPKTEESENLAAE \tabularnewline
& LLLVRAQEPITKMLEHEEKSVQVKTALKSKETKSEQHEAHIPVIIITTETGAALIEPKQLLYVPKIKTNLGACEGNDVRKEGVLRGLVLHRGTVLDLL \tabularnewline
& LETRTDQKKLETKTGHPKLPVSTQLEAPVIEAVLGKGLGHAEVHSQVLERRAELDELNEYILSGVAIPETMACELTVIIKTLVLQKKKVIRNIKLGLH \tabularnewline

\end{tabular}
}
\label{Seq_List_6}
\end{center}
\end{table}

\end{document}